\documentclass[reprint,sort&compress,superscriptaddress,amsmath,aps,showpacs,pra]{revtex4-1}
\usepackage{amssymb}
\usepackage{appendix}
\usepackage{verbatim}
\usepackage{float}
\usepackage{color}
\usepackage{textcomp}
\usepackage{gensymb}
\usepackage{hyperref}
\hypersetup{
     colorlinks   = true,
     citecolor    = blue
}
\usepackage{siunitx}
\usepackage{graphicx}
\usepackage{dcolumn}
\usepackage{bm}
\usepackage{braket}
\usepackage{empheq}
\usepackage{nicefrac} 
\usepackage[dvipsnames]{xcolor}
\renewcommand{\vec}[1]{\bm{#1}}

\RequirePackage{color}
\usepackage[normalem]{ulem}

\begin{document}

\title{Narrow bands, electrostatic interactions and band topology in graphene stacks} 

\author{Pierre A. Pantale\'on}
\email{ppantaleon@uabc.edu.mx}
\affiliation{Imdea Nanoscience, Faraday 9, 28015 Madrid, Spain}
\author{Tommaso Cea}
\affiliation{Imdea Nanoscience, Faraday 9, 28015 Madrid, Spain}
\author{Rory Brown}
\affiliation{Department of Physics and Astronomy, University of Manchester, Manchester, M13 9PY, UK} 
\author{Niels R. Walet}
\affiliation{Department of Physics and Astronomy, University of Manchester, Manchester, M13 9PY, UK} 
\author{Francisco Guinea}
\affiliation{Imdea Nanoscience, Faraday 9, 28015 Madrid, Spain} 
\affiliation{Donostia International Physics Center, Paseo Manuel de Lardiz\'abal 4, 20018 San Sebasti\'an, Spain}

\date{\today}

\begin{abstract}
The occurrence of superconducting and insulating phases is well-established in twisted graphene bilayers, and they have also been reported in other arrangements of graphene layers. We investigate three such arrangements: untwisted AB bilayer graphene on an hBN substrate, two graphene bilayers twisted  with respect to each other, and a single ABC stacked graphene trilayer on an hBN substrate. Narrow bands with different topology occur in all cases, producing a high density of states which enhances the role of interactions. We investigate the effect of the long range Coulomb interaction, treated within the self consistent Hartree-Fock approximation. We find that the on-site part of the Fock potential strongly modifies the band structure at charge neutrality. The Hartree part does not significantly modify the shape and width of the bands in the three cases considered here, in contrast to the effect that such a potential has in twisted bilayer graphene.  

\end{abstract}

\maketitle

\section{Introduction}

The recent discovery of superconductivity and insulating behavior in twisted graphene bilayers \cite{Cao2018,Cao2018_bis} and trilayers \cite{Park2021,Hao2021} has sparked  interest in other types of layered materials which may host narrow bands, built by stacking atomically thin two-dimensional layers. Observations of insulating behavior and/or superconductivity have been reported for graphene bilayers on hBN substrates \cite{Moriyama2019}, ABC-stacked trilayers on hBN substrates \cite{Chen2019,Chen2019a,Chen2020, Chittari2019a}, pairs of graphene bilayers twisted with respect to each other \cite{HeSym2020,Letal19,Cetal19,Tsai2019,Shen2019}, rhombohedral tetralayers \cite{Kerelsky2019b}, rhombohedral trilayer graphene~\cite{zhou2021half,zhou2021superconductivity} and twisted transition-metal-dichalcogenide layers \cite{Wang2019b}.
Theoretical calculations show that these systems host narrow bands, in some cases only upon the application of a perpendicular electric field \cite{HeSym2020,Tsai2019,Chen2020,Chen2019,Chen2020,Chen2019a,Chittari2019a,Kerelsky2019b,Wang2019b}. The width of these bands, $W \sim 5 - 15\,\text{meV}$, can be as small as that for twisted graphene bilayers near magic angles \cite{Bistritzer2011}.

The bands of twisted bilayer graphene (TBG) show a number of unique properties not commonly found in other crystalline materials with narrow bands: i) the low Fermi velocity is not caused by large self-energy corrections, as in many strongly correlated materials~\cite{D77,S84,S11,KKNUZ15}, ii) the Wannier functions extend throughout the unit cell, and different Wannier functions are located in the same region of space \cite{Po2018,Koshino2018a,Kang2018b,Angeli2018a}, iii) the wavefunctions at different points in the Brillouin Zone differ significantly \cite{Rademaker2018a}, and iv) electronic interactions tend to increase the bandwidth (see the theoretical calculations \cite{Guinea2018a,Xie2020a,Liu2019a,Cea2019,Rademaker2019a,Cea2019} and related experimental results \cite{Kerelsky2019a,Xie2019,Jiang2019a,Choi2019,Wong2019,Tsai2019,Zondiner2019}). 

In the following, we analyze whether the narrow bands found in other graphene stacks share some of the features of twisted bilayer graphene. We focus on the role of the long-range electrostatic interactions whose strength is of the order of $e^2 / \epsilon L$, where $\epsilon$ is the dielectric constant of the surrounding medium and $L$ the length of the Moir\'e unit cell; this is comparable to or larger than the bandwidth. We describe this interaction by means of a self-consistent Hartree-Fock approximation. 
This variational technique gives a reasonable description of many interacting quantum systems. For twisted bilayer graphene it describes both the change in the width and shape of the central bands, as well as the existence of different broken symmetry phases~\cite{Guinea2018a,Cea2019,Cea2020c}

The paper is organized as follows. In Sec.~\ref{sec: The Models} we present the low energy models used. The band structure in the presence of long-range electrostatic interactions and the charge density of: AB bilayer graphene on hBN, a single ABC stacked graphene trilayer on hBN and two graphene bilayers twisted with respect to each other are presented in  Sec.~\ref{sec: Graphene bilayer aligned with hBN}, Sec.~\ref{sec: ABC graphene trilayer aligned with hBN} and Sec.~\ref{sec: Two twisted graphene bilayers}, respectively. An analysis of the Fock interactions is presented in Sec.~\ref{sec: Fock Interaction}. Finally, conclusions are given in Sec.~\ref{sec: Conclusions}.

 \begin{figure*}       
\begin{centering}
\includegraphics[scale=0.29]{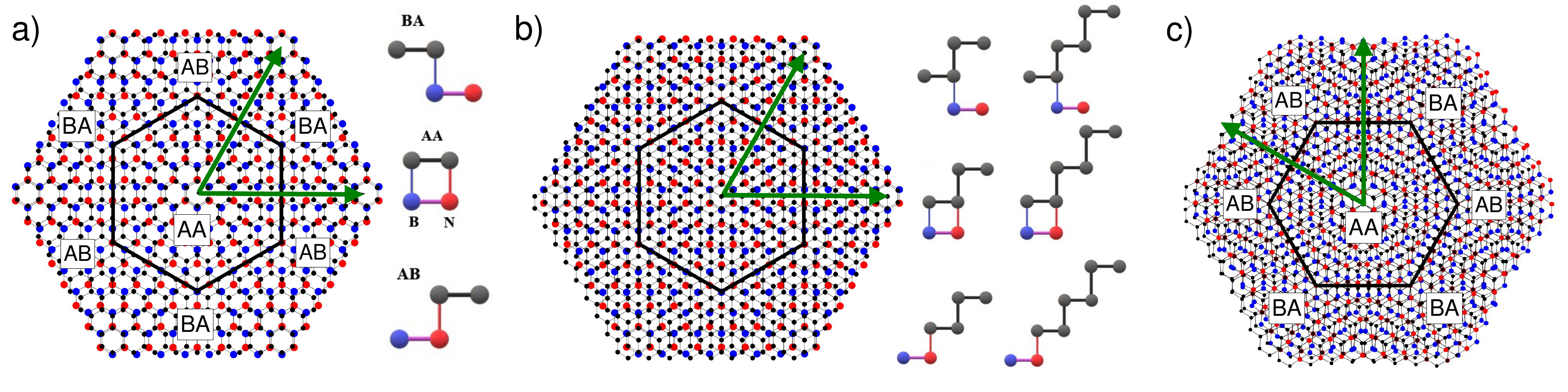}
\par\end{centering}
\caption{a) Representation of a graphene/hBN moir\'e superlattice and the stacking arrangements appearing in the three configurations $BA$, $AA$ and $AB$. Black points represent carbon atoms, blue boron and red nitrogen. In the figure, we exaggerate the lattice mismatch ($\sim 14\%$). In real systems, the lattice constants have a mismatch $\sim 1.8\%$ which gives rise to a moiré superlattice of size $L\approx 15 \text{ nm}$. b) In bilayer graphene on hBN (BG/hBN) the second graphene layer is $AB$ aligned to the first. In this case, as in ABC trilayer graphene on hBN (TG/hBN) the superlattice potential is acting only over the closest graphene layer. As in a) the stacking arrangement for both BG/hBN and TG/hBN is shown. c) Representation of two twisted graphene bilayers (TDBG) with $ABA'B'$ stacking. In the figure, to visualize the moir\'e pattern, we set a twist angle $\theta\sim 7^\circ$. Labels indicate the stack between sites of the two middle layers. In all figures, green arrows are the superlattice moir\'e vectors and the black hexagon represents the moir\'e unit cell.
\label{fig: FigRealSpace}}
\end{figure*}
 
 \begin{figure}[ht]       
\begin{centering}
\includegraphics[scale=0.17]{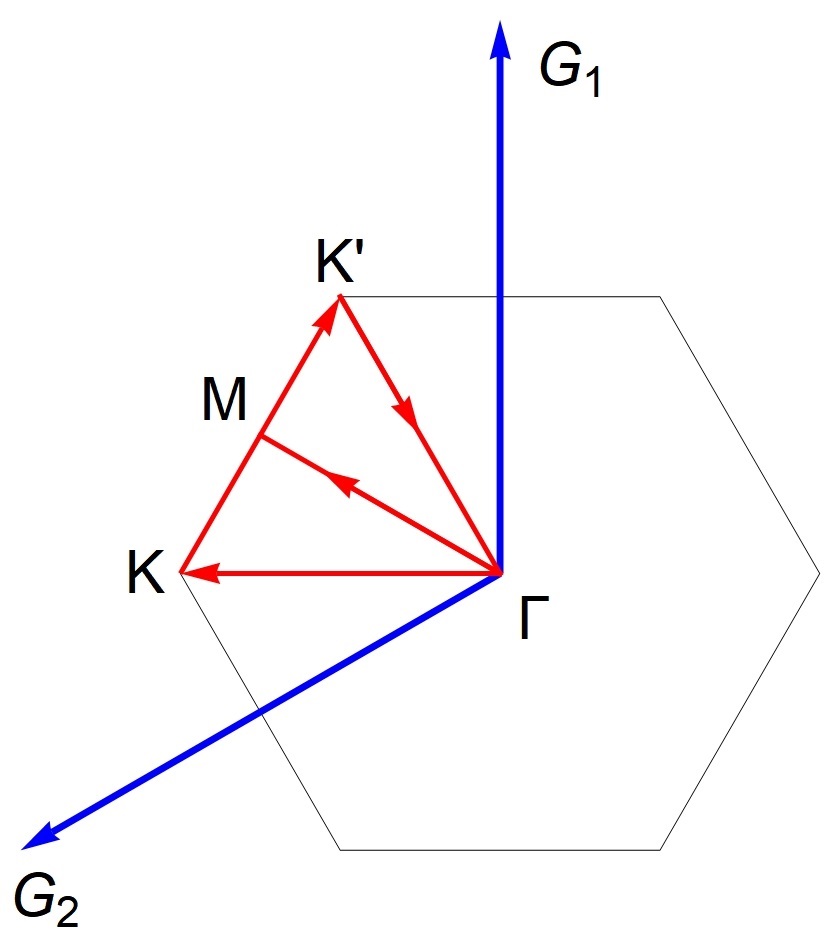}
\par\end{centering}
\caption{BG/hBN and TG/hBN moir\'e Brillouin zone. $G_{1}$ and $G_{2}$ are the reciprocal lattice vectors. The red circuit through the Brillouin zone, $\vec \Gamma \vec K \vec K' \vec \Gamma \vec M $, is used to display the band structure. 
\label{fig: FigbZ}}
\end{figure}
 
 \begin{figure*}     
\begin{centering}
\includegraphics[scale=0.55]{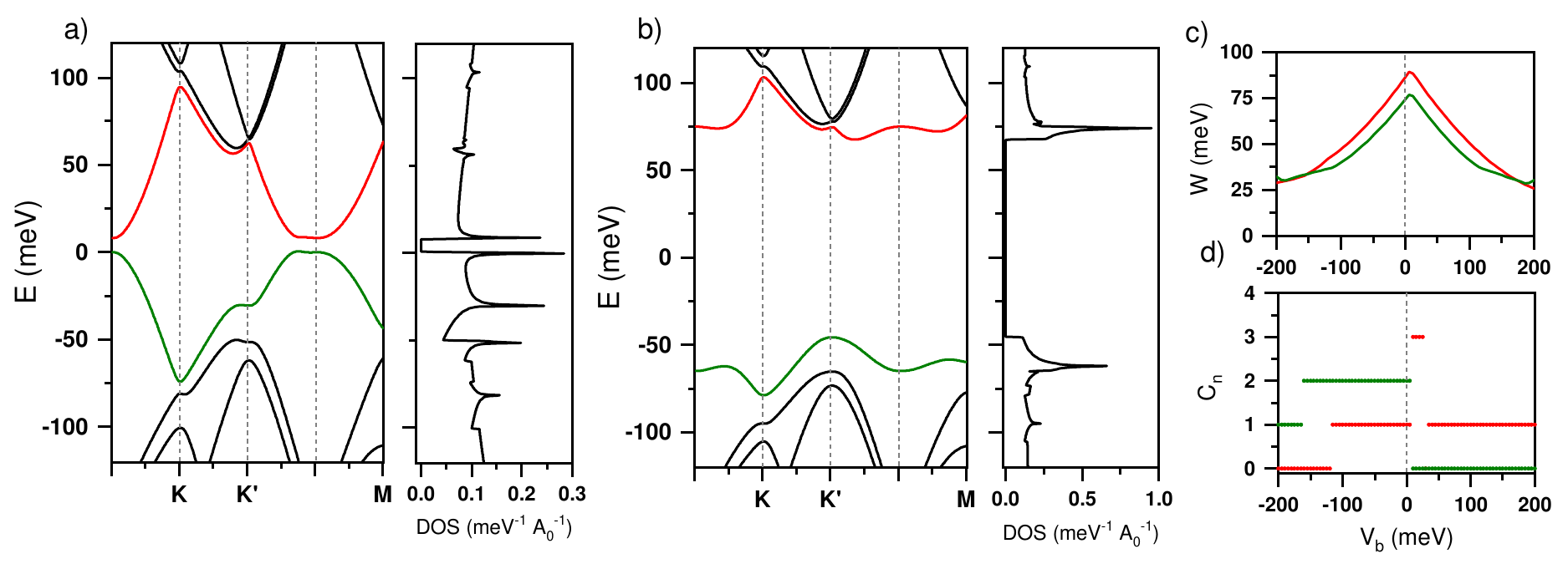} 
\par\end{centering}
\caption{Band structure of AB bilayer graphene mounted on hBN. a) Case $V_{b}=0$: there are two bands (red and green lines for the lower and upper band, respectively) near the charge neutrality point. b) Case $V_{b}=150$ meV. We see that if the displacement field is increased the bandwidth of the middle bands is reduced. In c) and d) we show the bandwidth and the Chern numbers of the middle bands as a function of the displacement field, respectively. The two middle bands are consistently labelled by the same color in all figures.}  
\label{fig: FigBGbands}
\end{figure*}

\begin{figure}       
\begin{centering}
\includegraphics[scale=0.57]{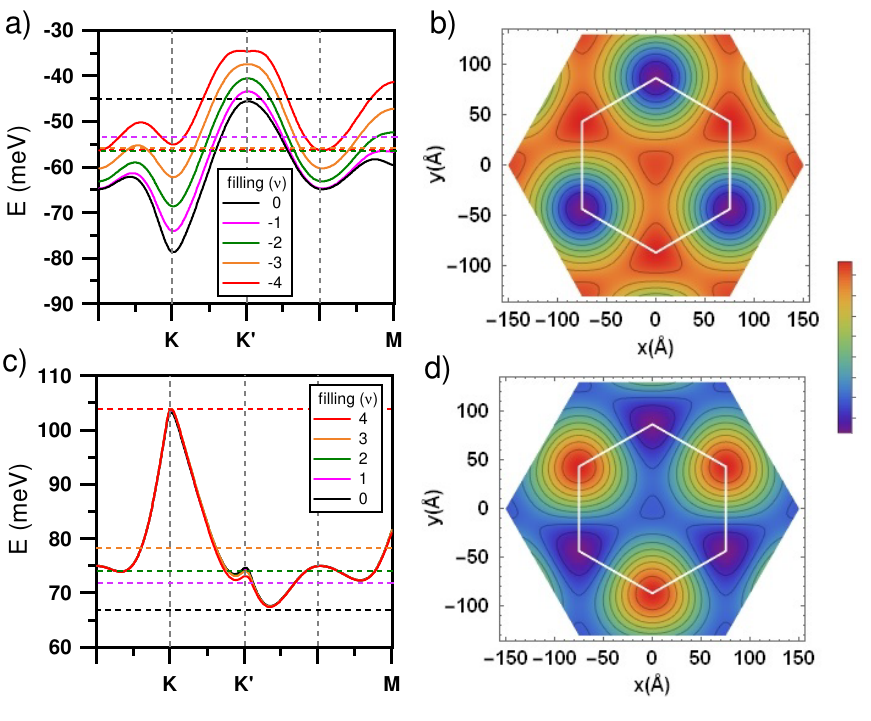}
\par\end{centering}
\caption{Self-consistent bands for BG/hBN in the presence of a Hartree potential and the same parameters as in Fig.~\ref{fig: FigBGbands}b). In a) and c) we plot the lower and upper narrow band with negative and positive filling fractions, respectively. Panels b) and d) display the real space distribution of the Hartree potential for a filling fraction  $\nu=-2$ and $\nu=2$, respectively. The white hexagon shows the real-space unit cell. The horizontal dashed lines in a) and c) are the corresponding Fermi energies.
\label{fig: FigBilHar4}}
\end{figure}

\section{The Models} \label{sec: The Models}
 We analyze two distinct types of systems: i) bilayers and trilayers without a twist, where the flat bands are induced by a combination of a strain due to the substrate and a perpendicular electric field, and ii) a pair of graphene bilayers which are twisted by a small angle relative to each other. In both cases we analyze the electronic bands by means of continuum models, based on the Dirac equation for a single graphene layer (see Appendix A).  The first case, which involves bilayers and trilayers, requires models for the interlayer hopping in the multilayer and for the interaction between the substrate and the multilayer. 
 
 We model the stacks of graphene on hBN with a low-energy continuum model. The hBN substrate is modelled in the low-energy Hamiltonian as a modulated effective field affecting only the nearest graphene layer. As is shown in Fig.~\ref{fig: FigRealSpace}, the substrate induces a triangular superlattice with periodicity $L \sim 15$ nm and reciprocal lattice vectors  $\vec{G}_1 =4\pi \left(0,1\right)/L$ and $\vec{G}_2 =2\pi \left(-\sqrt{3},-1\right)/L$. The reciprocal lattice has a reduced superlattice or moir\'e Brillouin zone (mBZ), with new symmetry points $\boldsymbol{K=}\frac{1}{3}\left(G_{1}+2G_{2}\right)$ and $\boldsymbol{K'=}\frac{1}{3}\left(2G_{1}+G_{2}\right)$. The area of the mBZ is a factor $L^2$ smaller than that of graphene: we can therefore perform zone folding to study the system, decomposing the momentum $\vec{g}_{mn}$ within the graphene Brillouin zone into a momentum $\vec{k}$ within the boundaries of the mBZ and a contribution from repeats of the superlattice mBZ. We define the momentum $\vec{k}$ inside the mBZ such that
\begin{equation}
\vec{g}_{mn} = \vec{k} + (m,n)\cdot (\vec{G}_1, \vec{G}_2) \equiv \vec{k} + \vec{G}_{mn},
\label{eq:Gmn}
\end{equation}
where $\vec{G}_{mn}=m\vec{G}_1+n\vec{G}_2$ with $m,n$ integers. Each $\vec{G}_{mn}$ vector in the reciprocal space has six nearest neighbors, at a displacement $\vec{G}_i=(4\pi/\sqrt{3}L)\left(\cos\frac{2j+1}{6}\pi,\sin\frac{2j+1}{6}\pi\right), i= 1,2,...,6$; these generate the first harmonic functions of the superlattice. We refer to these vectors as the ``first star" of reciprocal lattice vectors.

It has been established \cite{Wallbank2013a, San-Jose2014, Mucha-Kruczynski2013a,Jung2017} that a minimal valley-symmetric model for the low-energy electronic structure of graphene/hBN is given by a Hamiltonian of the form
\begin{equation}
H=H_{0}+V_\text{SL}+V_\text{H}+V_\text{F},\label{eq: Main Hamiltonian}
\end{equation}
where the first term represents the mono or multilayer graphene low-energy Hamiltonian in the vicinity of a single valley, expanded about the $K$ point. The second term $V_\text{SL}$ describes the effect of the hBN on the closest graphene layer, treating the resultant superlattice as a perturbation on the Dirac Hamiltonian. To take into account the long-range Coulomb interaction we also introduce self-consistent Hartree and Fock potentials, $V_\text{H}$ and $V_\text{F}$ respectively (see section VI and appendix B for further details). The effect of a hBN substrate on charge carriers in a graphene monolayer can be described by including the first harmonic functions of the superlattice. Including the interactions between charge carriers in graphene and the hBN substrate as a perturbing potential $V_{SL}$, the Hamiltonian in Eq.~\eqref{eq: Main Hamiltonian} can be written as
\begin{align}
H & = H_0(\vec{g}_{mn})\otimes \mathbb{I}_N + V_{SL} + V_\text{H}+V_\text{F}  \notag \\
& = H_0(\vec{k})\otimes \mathbb{I}_N + H_0(\vec{G}_{mn})\otimes \mathbb{I}_N + V_{SL} + V_\text{H}+V_\text{F} \notag \\
& = H_0(\vec{k})\otimes \mathbb{I}_N + H_{SL} + V_\text{H}+V_\text{F},
\label{eq: hamprincipal}
\end{align}
where $H_{SL}$ contains all superlattice effects, $H_0(\vec{k})$ is the isolated mono or multilayer graphene Hamiltonian in the mBZ and $N$ the number of reciprocal lattice vectors in the Fourier expansion. In BG/hBN  and TG/hBN we can assume that the superlattice potential $V_{SL}$ affects only the closest graphene layer and couples each momentum ${G}_{mn}$ with their six nearest neighbors, $G_{j}$. The mismatch between the lattice constants of graphene and hBN leads, in the aligned situation, to the formation of a triangular moir\'e superlattice where the length of the unit vector is $L \approx 15$ nm. We describe the effect of hBN on the nearest graphene layer by an effective potential, which is written as a Fourier expansion whose components are labeled by reciprocal lattice vectors of the superlattice~\cite{Wallbank2013a}. We assume that these components decay rapidly, and keep only the ones associated to the first star of  reciprocal lattice  vectors.  The effective potential periodic in the moir\'e unit cell is~\cite{Wallbank2013a},
\begin{equation}
V_\text{SL}\left(\vec{r}\right)=
w_{0}\sigma_{0}+\Delta\sigma_{z}+
\sum_{j}
v_\text{SL}(\vec{G}_j)e^{i\vec{G}_j\cdot\vec{r}},
\end{equation}
with amplitudes $v_\text{SL}(\vec{G}_j)$ given by
\begin{equation}
v_\text{SL}(\vec{G}_j)= V_{s}(\vec{G}_j)+V_{\Delta}(\vec{G}_j)+V_{g}(\vec{G}_j),
\label{eq: hbN Perturbation}  
\end{equation}
where
\begin{align}
V_{s}(\vec{G}_j)&= \left[V_{s}^{e}+i(-1)^{j}V_{s}^{o}\right]\sigma_{0},\\ \nonumber V_{\Delta}(\vec{G}_j)&=\left[V_{\Delta}^{o}+i(-1)^{j}V_{\Delta}^{e}\right]\sigma_{3}, \\ \nonumber
V_{g}(\vec{G}_j)&=\left[V_{g}^{e}+i(-1)^{j}V_{g}^{o}\right]M_j,\\ \nonumber
\end{align}
with $M_j=(-i\sigma_{2}G_{j}^{x}+i\sigma_{1}G_{j}^{y})/|G_{j}|$. The $2 \times 2$ Pauli matrices act on the sublattice index in a single graphene layer. The parameters $w_{0}$ and $\Delta$ represent a spatially uniform scalar and mass term (note that hBN breaks inversion symmetry, and allows for a mass term~\cite{Hunt2013}). The parameters $V_{s}^{e}$ and $V_{s}^{o}$ are position-dependent scalar terms and are respectively even and odd under spatial inversion. Similarly, $V_{\Delta}^{o(e)}$ and $V_{g}^{o(e)}$ are position dependent mass and gauge terms.  As shown in Fig.~\ref{fig: FigRealSpace}b), there are three different configurations with different sets of parameters depending on the stack configuration at the unit cell origin. In TBG on hBN these are nonequivalent, resulting in different band structures \cite{Cea2020a}. However, in both BG/hBN and TG/hBN the different stack configurations generate the same band structure. In particular, we use the set of parameters for the $AA$ configuration given by:
$(w_{0},\Delta,V_{s}^{e},V_{s}^{o},V_{\Delta}^{e},V_{\Delta}^{o},V_{g}^{e},V_{g}^{o})= (0, 3.6, -1.88, 6.78, 0.017, -6.85, 3.61, -12.4)\text{ meV}$. The configurations $AB$ and $BA$ are related to the $AA$ configuration by a rotation of $\pm 2\pi/3$ in the parameter space, as detailed in Ref.~\cite{Jung2017}.

\begin{figure*}    
\begin{centering}
\includegraphics[scale=0.37]{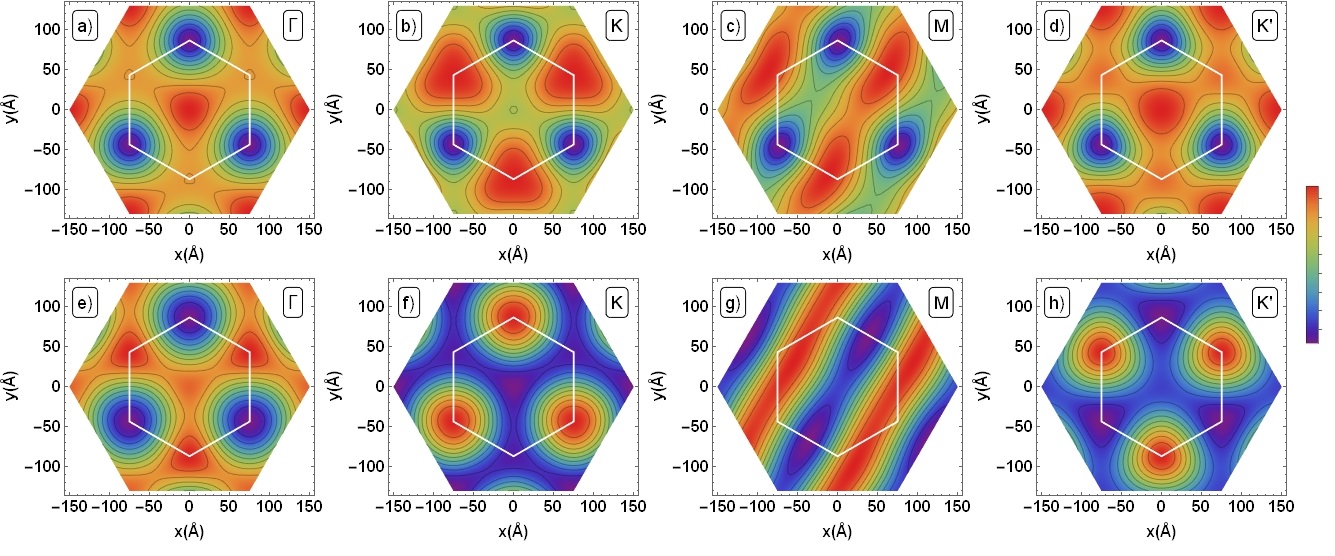}
\par\end{centering}
\caption{Charge density at the high symmetry points within the mBZ for the bands of BG/hBN in Fig.~\ref{fig: FigBilHar4} with a filling fraction $\nu=-2$ (top row) and $\nu=2$ (bottom row). The electric bias is set to $V_{b}=150$ meV. The corresponding symmetry point is indicated in each panel. The scale ranges from purple at the minimum to red for the maximum charge density. 
\label{fig: FigBGDens}}
\end{figure*}

\begin{figure*}     
\begin{centering}
\includegraphics[scale=0.55]{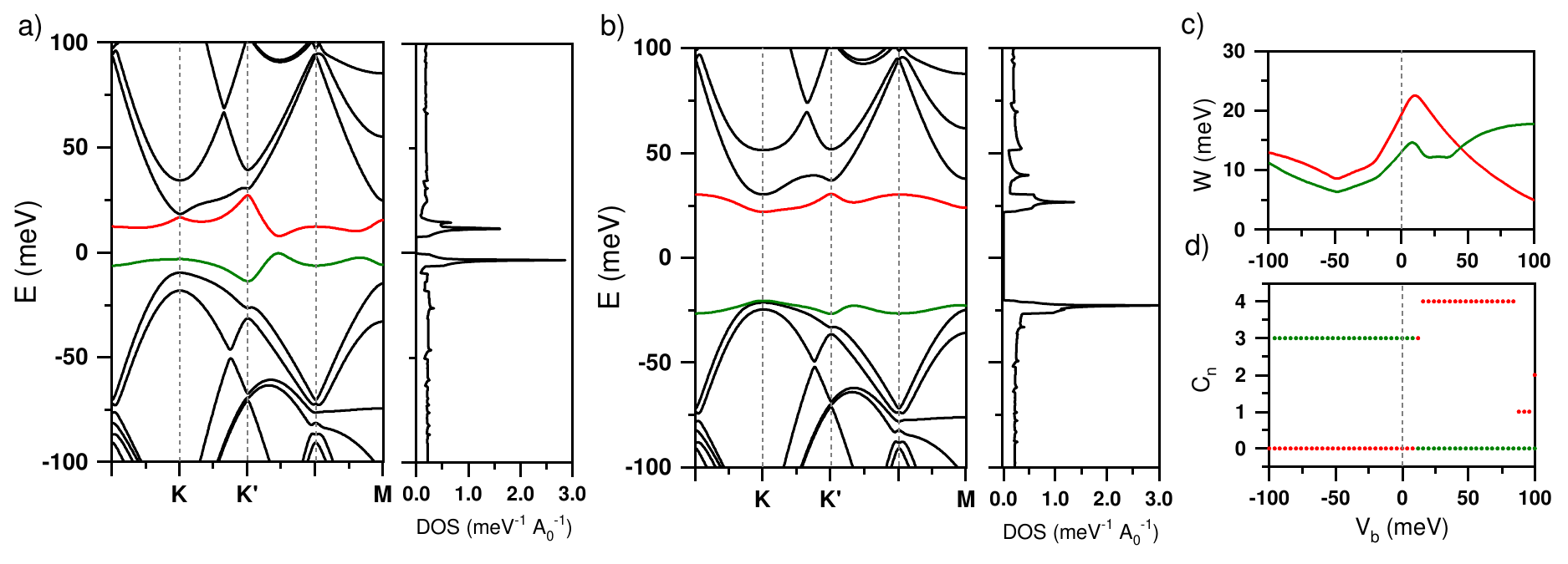} 
\par\end{centering}
\caption{Band structure of ABC trilayer graphene mounted on hBN. a) $V_{b}=0$. There are two narrow bands with a large density of states (red and green lines for the lower and upper band, respectively). b)  $V_{b}=-48$ meV. When the electrostatic potential is modified the bandwidth of the middle bands is reduced. In c) and d) we show the bandwidth and the Chern number of the middle bands as a function of the electrostatic potential, respectively. The color scheme of the middle bands is the same in all figures.  
\label{fig: FigTGbands}}
\end{figure*}

In order to carry out the mean-field calculations, the charge density induced by the occupied states needs to be calculated. We assume approximate local charge neutrality at half filling. Deviations from charge neutrality are due to contributions from electrons or holes induced by an external gate. The charge density is expanded in Fourier components defined by the moir\'e reciprocal lattice vectors. The charge density includes a constant term, which defines the average charge in the system. The electrostatic potential induced by this term is cancelled by the gate potential. In twisted bilayer graphene the charge density converges rapidly: only six equivalent reciprocal lattice vectors need to be considered \cite{Guinea2018a,Cea2019}, resulting in real components of the Hartree matrix elements. The charge density in the systems described here is more complex and requires additional Fourier components to achieve convergence, as the Hartree matrix elements contain both real and imaginary components and in some cases depend on the filling fraction. We will now outline each system in turn.

\section{Graphene bilayer aligned with hBN} \label{sec: Graphene bilayer aligned with hBN}

\subsection{Band structure}

We first consider bilayer graphene aligned with hBN. Using the standard parametrization for the hoppings between next nearest layers based on the SWM mode for graphite~\cite{McCann2013}, the bilayer Hamiltonian
 \begin{equation}
H_{0}=\left(\begin{array}{cccc}
-\frac{V_{b}}{2} & v_{0}\pi^{\ast} & -v_{4}\pi^{\ast} & v_{3}\pi\\
v_{0}\pi & -\frac{V_{b}}{2}+\varDelta^{\prime} & \gamma_{1} & -v_{4}\pi^{\ast}\\
-v_{4}\pi & \gamma_{1} & \frac{V_{b}}{2}+\varDelta^{\prime} & v_{0}\pi^{\ast}\\
v_{3}\pi^{\ast} & -v_{4}\pi & v_{0}\pi & \frac{V_{b}}{2}
\end{array}\right)
\label{eq: BilMatrix},
\end{equation} 
is used in Eq. \ref{eq: Main Hamiltonian}. Here $\pi=\hbar k_{x}+i \hbar k_{y}$, the parameters $v_{i}=(\sqrt{3}/2)\gamma_i a/\hbar$ have dimensions of velocity, the lattice constant of graphene $a=0.246$~nm, $\gamma_{1}$ is the interlayer coupling and $\varDelta^{\prime}$ is
the energy difference between dimer and non-dimer sites~\cite{McCann2013}. The parameter values in Eq.~\eqref{eq: BilMatrix} are: $\gamma_{0}=3.16$,  $\gamma_{1}=0.381$, $\gamma_{3}=0.38$, $\gamma_{4}=-0.14$ and $\Delta'=0.022$~eV~\cite{Kuzmenko2009a}. We also introduce a perpendicular displacement field $V_{b}$ between graphene layers.

As shown in Fig.~\ref{fig: FigRealSpace}a), the arrangement between graphene and hBN allows for three stack configurations depending on the origin of the moir\'e unit cell. In BG/hBN they are equivalent if only a single graphene layer is aligned with hBN~\cite{Jung2017}. The band structure for BG/hBN is shown in Fig.~\ref{fig: FigBGbands}a). The substrate induces two middle bands, with a large density of states near charge neutrality. The gap between these bands has a value of $\Delta_{g}\sim 8$~meV, and can be tuned by the displacement field as shown in Fig.~\ref{fig: FigBGbands}c). As shown in Fig.~\ref{fig: FigBGbands}b) and Fig.~\ref{fig: FigBGbands}c), the bandwidth of these bands is sensitive to the displacement field and decreases almost linearly with the field magnitude. Our results are in agreement with previous DFT calculations in a similar system~\cite{Ramasubramaniam2011} .

The presence of hBN breaks both particle-hole and inversion symmetry. Since both middle bands are isolated for a range of parameters, the broken inversion symmetry allow us to calculate the band topology. Figure~\ref{fig: FigBGbands}d) displays the Chern number of a single valley as a function of the displacement field. Because of the time reversal symmetry, the Chern numbers of the opposite valley are opposite in sign. This system possesses different topological phases which can be tuned by the displacement field, and as shown in the appendix D, these phases also depend on the long-range Coulomb interaction.

\begin{figure} 
\begin{centering}
\includegraphics[scale=0.60]{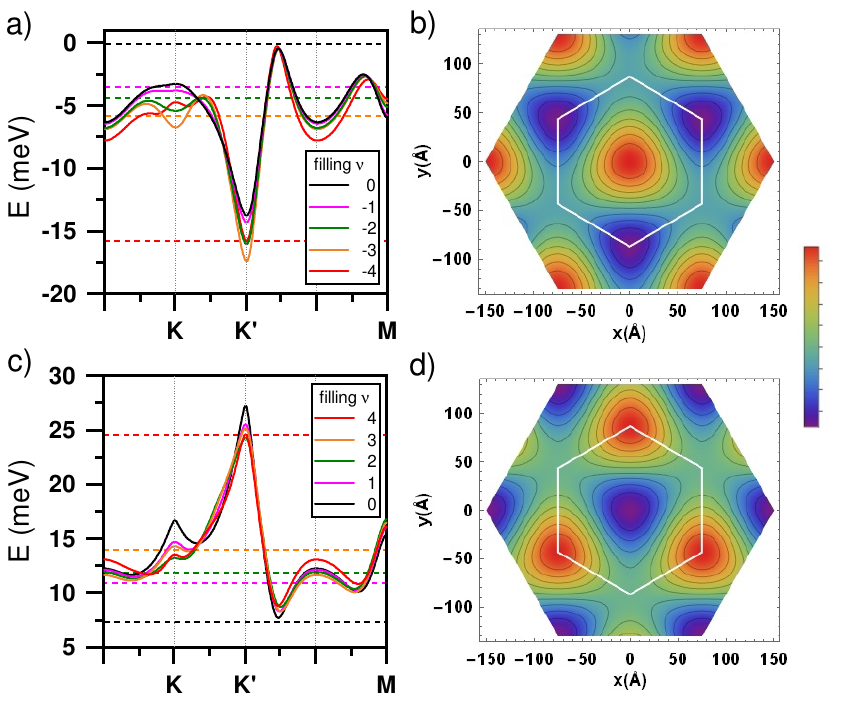}
\par\end{centering}
\caption{Self-consistent bands for TG/hBN in the presence of a Hartree potential and the same parameters as in Fig.~\ref{fig: FigTGbands}a). In a) and c) we plot the lower and upper narrow band with negative and positive filling fractions, respectively. The real space distribution of the Hartree potential is shown for a filling b) $\nu=-2$ and d) for $\nu=2$. The horizontal dashed lines in a) and c) are the Fermi energies corresponding to the filling fraction denoted by that color. 
\label{fig: FigTrilHar4}}
\end{figure}

\begin{figure*}     
\begin{centering}
\includegraphics[scale=0.30]{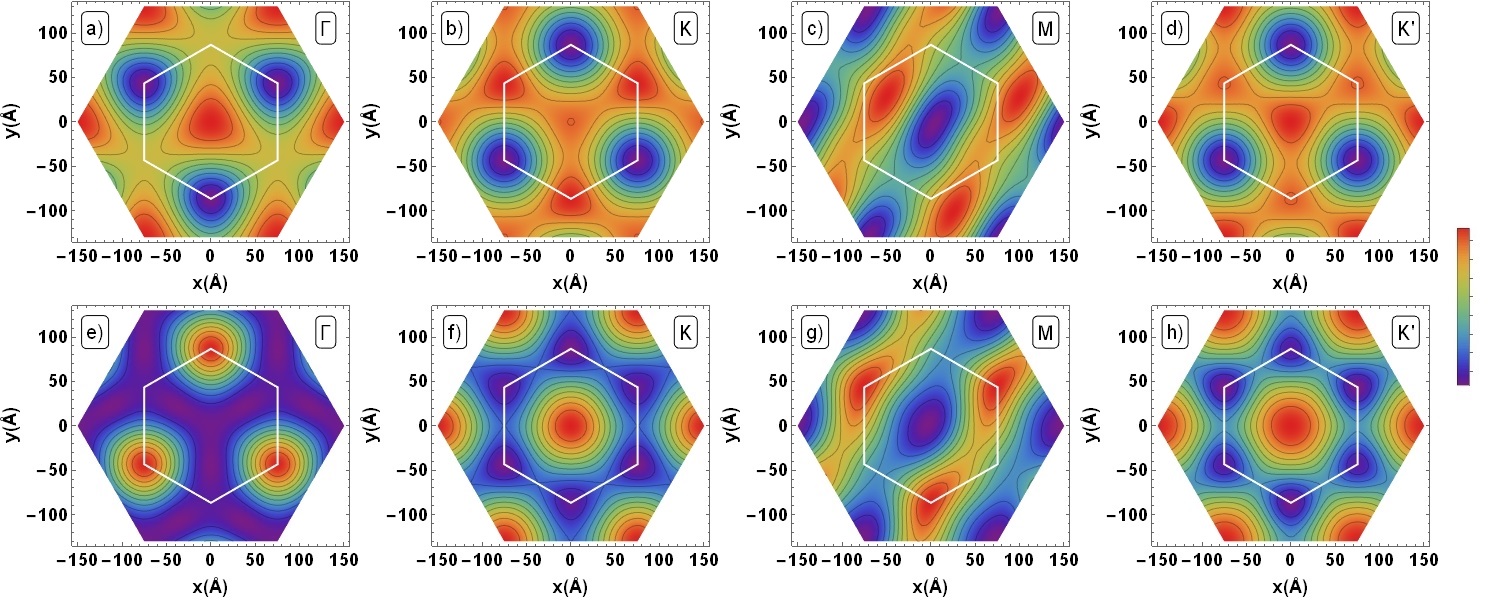}
\par\end{centering}
\centering{}\caption{Charge densities at the high symmetry points within the mBZ for the bands of TG/hBN in Fig.~\ref{fig: FigTrilHar4} with a filling fraction $\nu=-2$ (top row) and $\nu=2$ (bottom row). The electric bias is set to $V_{b}=0$ meV. The corresponding symmetry point is indicated in each panel. The scale ranges from purple at the minimum to red for the maximum charge density. \label{fig: FigTGDens}}
\end{figure*}

\subsection{Self-consistent Hartree interaction}

Recently Moriyama \emph{et al.} \cite{Moriyama2019} reported signatures of fragile superconductivity in a hBN/BG/hBN system, and Zheng \emph{et al.} \cite{Zheng2020uf} found unconventional ferroelectricity in a similar configuration. These results indicate that there are correlated effects due to the presence of middle narrow bands under certain conditions. As shown in Fig.~\ref{fig: FigBGbands}c), for $V_{b}=0$ the two middle bands have bandwidths of $74$ and $86$ meV respectively. The bandwidth is reduced by increasing the magnitude of the displacement field. In the presence of a long-range Coulomb interaction, the inhomogeneous charge distribution leads to an electrostatic potential of $V_{0}=e^2/\epsilon L$, with $L$ the moir\'e length and $\epsilon$ the dielectric constant. For a small screening of $\epsilon=4$, $V_{0}\sim 24$ meV. For a large value of $V_{b}$ the bandwidth of the narrow bands becomes comparable with the effective electrostatic potential, as seen in Fig.~\ref{fig: FigBGbands}d). 

We consider the effect of a self-consistent Hartree interaction, where the self consistent bands for different fillings and the distribution of the Hartree potential in real space (Eq.~\eqref{eq: VhRealSpace} in appendix B) are shown in Fig.~\ref{fig: FigBilHar4}. We focus on the case with $V_{b}= 150$ meV where the bands are narrow. As shown in Fig.~\ref{fig: FigBGDens}, these charge densities are significantly different from those in twisted bilayer graphene near magic angles~\cite{Rademaker2018a,Guinea2018a}. As discussed in Refs.~\cite{Guinea2018a,Cea2019}, in TBG the dominant contribution to the Hartree interaction comes from the first star of reciprocal lattice vectors. In BG/hBN up to three stars are required to achieve convergence. The difference in the Hartree potential strength of this system and TBG is due to the different distribution of the wavefunctions in real space. In TBG the wavefunctions are centered at the middle AA stacked region, or in a ring around it~\cite{Rademaker2018a}. In the Fourier expansion of the Hartree potential, only a single star with real components of the charge density was required \cite{Guinea2018a}. As shown in Fig.~\ref{fig: FigRealSpace}a), in BG/hBN the substrate has different atoms in each sublattice with different stack configurations, and the Hartree interaction contains both symmetric and antisymmetric components of the charge density. 

Figure.~\ref{fig: FigBGDens} shows the charge density in real space for different points within the mBZ. The individual figures correspond to the high symmetry points for the narrow bands in Fig.~\ref{fig: FigBilHar4}a) and Fig.~\ref{fig: FigBilHar4}c) with filling $\nu=-2$ and $\nu=2$ on the top and bottom row, respectively. The white hexagon is the real space unit cell. For a negative filling fraction $\nu=-2$ we find that all the states in the mBZ (top row in Fig.~\ref{fig: FigBGDens}), have density distribution that is high where the Hartree potential peaks [Fig.~\ref{fig: FigBilHar4}b)]. As a consequence the bands are uniformly shifted, as shown in Fig.~\ref{fig: FigBilHar4}a). However,  for a positive filling $\nu=2$ only the states near $\vec K'$ [Fig.~\ref{fig: FigBGDens}h)] have similar sensity distributions as the Hartree potential density [Fig.~\ref{fig: FigBilHar4}d)], and thus only these states are sensitive to the Hartree interaction. This is shown in Fig.~\ref{fig: FigBilHar4}c) where there is a small distortion around the $\vec K'$ point. 
The other regions are insensitive to the Hartree potential. We find the same behaviour for different fillings fractions and different displacement fields. In twisted bilayer graphene, the Hartree potential is strong and modifies the shape of the band, increasing its width to the order of magnitude of $V_{0}$. In the present case even for a relatively low dielectric constant ($\epsilon = 4$) the bands are not significantly deformed, as seen in Fig.~\ref{fig: FigBilHar4}. These results indicate that the effect of the Hartree potential is significantly smaller in this case when compared to twisted graphene bilayers~\cite{Guinea2018a,Cea2019}

\section{ABC graphene trilayer aligned with hBN} \label{sec: ABC graphene trilayer aligned with hBN}

\subsection{Band structure}

We now consider ABC trilayer graphene aligned with hBN, with a Hamiltonian $H_{0}$ in Eq.~\eqref{eq: Main Hamiltonian} given by~\cite{Zhang2010}
\begin{equation}
H_{0}=\left(\begin{array}{cccccc}
-\frac{V_{b}}{2}+\delta & v_{0}\pi^{\dagger} & v_{4}\pi^{\dagger} & v_{3}\pi & 0 & \frac{1}{2}\gamma_{2}\\
v_{0}\pi & -\frac{V_{b}}{2} & \gamma_{1} & v_{4}\pi^{\dagger} & 0 & 0\\
v_{4}\pi & \gamma_{1} & 0 & v_{0}\pi^{\dagger} & v_{4}\pi^{\dagger} & v_{3}\pi\\
v_{3}\pi^{\dagger} & v_{4}\pi & v_{0}\pi & 0 & \gamma_{1} & v_{4}\pi^{\dagger}\\
0 & 0 & v_{4}\pi & \gamma_{1} & \frac{V_{b}}{2} & v_{0}\pi^{\dagger}\\
\frac{1}{2}\gamma_{2} & 0 & v_{3}\pi^{\dagger} & v_{4}\pi & v_{0}\pi & \frac{V_{b}}{2}+\delta
\end{array}\right)\,,
\label{eq:trilMatrix} 
\end{equation} 
where we use the parameters derived from DFT calculations given in Ref.~\cite{Zhang2010}: $\gamma_{0}=3.16$, $\gamma_{1}=0.502$, $\gamma_{2}=-0.0171$, $\gamma_{3}=-0.377$, $\gamma_{4}=-0.099$, $\delta=-0.0014\text{ eV}$. As before, the terms $\gamma_{0}$ and $\gamma_{1}$ are the nearest-neighbor inter- and intralayer hopping parameters. The term $\gamma_{2}$ is the hopping parameter between the first and third layer. The interlayer potential $\delta$ accounts for a possible difference in the on-site energies between the first and third layer.
It has recently been shown that the presence of the superlattice potential induced by hBN in TG gives rise to a tunable Mott insulating behavior and reveals signatures of superconductivity \cite{Chen2019,Chen2019a}. This tunability is introduced by applying a perpendicular displacement field~\cite{Lui2011}. As shown in Fig.\ref{fig: FigRealSpace}b), there are three distinct stack configurations between trilayer graphene which are equivalent if the hBN is aligned only with the lower graphene layer. In the following, we consider the same superlattice parameters as in the previous case.  

Figure~\ref{fig: FigTGbands}a) shows the band structure for TG/hBN, where the superlattice potential induces two narrow bands with a large density of states near charge neutrality. We find that the gap between these bands $\Delta_{g}\sim~8.6$~meV with bandwidths of $13$ and $19$ meV. As shown in Fig.~\ref{fig: FigTGbands}c) both gap and bandwidth are tunable by a perpendicular displacement field. For example, the bandwidth of the lower band can be reduced by up to $6$ meV as shown in Fig.~\ref{fig: FigTGbands}b). The results for the bands in the absence of interactions are in reasonable agreement with those reported in Refs.~\cite{Chen2019,Chen2020,Chittari2019a}, as seen in Fig.~\ref{fig: FigTGbands}. The presence of the hBN breaks both particle-hole and inversion symmetry, and since both bands are well separated from the others we can calculate their Chern numbers. Figure.~\ref{fig: FigTGbands}d) shows the Chern number as a function of the displacement field. This system possesses different topological phases which strongly depend on the displacement field, in agreement with Ref.~\cite{Galeano2021}. 

\subsection{Self-consistent Hartree interaction}

It has been recently shown that TG/hBN is a gate tunable Mott~\cite{Chen2019a} and Chern insulator~\cite{Chen2020} and may support spin/valley polarization~\cite{Galeano2021}, ferromagnetism~\cite{ZhangTGhBN2019,RepellinFer2020,Chen2020, zhou2021half} and superconductivity~\cite{zhou2021superconductivity}. As shown in Fig.~\ref{fig: FigTGbands}, the moir\'e superlattice of TG/hBN hosts narrow bands for different values of the displacement field. The bandwidth of the narrow bands is comparable to and somewhat smaller than the Coulomb interaction strength for all values of the displacement field, as shown in Fig.~\ref{fig: FigTGbands}c). In particular, for a screening of $\epsilon =  4$ the effective Coulomb interaction $V_{0}=24$ meV is larger than the bandwidth of both narrow bands in a large interval of the displacement field. 

Similar to the previous case, we find that the comparison of the bandwidth with the effective Coulomb potential does not ensure an strong Hartree interaction. As shown in Fig.~\ref{fig: FigTrilHar4} for $\epsilon = 4$, the Hartree effects weakly modify the width and shape of the bands. We consider the case of $V_{b}=0$, but similar results are obtained for different displacement field values. 
Similar to BG/hBN, the Fourier expansion of the Hartree potential contains both symmetric and antisymmetric components of the charge density. This is shown in Fig.~\ref{fig: FigTrilHar4}b) and d) where we display the real space distribution of the Hartree potential, (Eq.~\eqref{eq: VhRealSpace} in appendix B),  at different filling fractions.

Similar to the case of BG/hBN, the small distortions in the band structure are due to the different distribution of the charge density at the different points within the mBZ. In the case of $\nu=-2$ shown in Fig.~\ref{fig: FigTrilHar4}a) there is a small distortion near $\vec K$ and a band shifting at $\vec K'$. The effective Hartree potential is shown in Fig.~\ref{fig: FigTrilHar4}b).
As a function of filling, the charge distribution is non-uniform and its net contribution to the Hartree potential is small. We find the same behaviour for different fillings and different displacement fields. These results imply that the Hartree interactions in TG/hBN (and BG/hBN) are significantly smaller than those obtained for twisted graphene bilayers~\cite{Guinea2018a,Cea2019}. 

\begin{figure*}     
\begin{centering}
\includegraphics[scale=0.50]{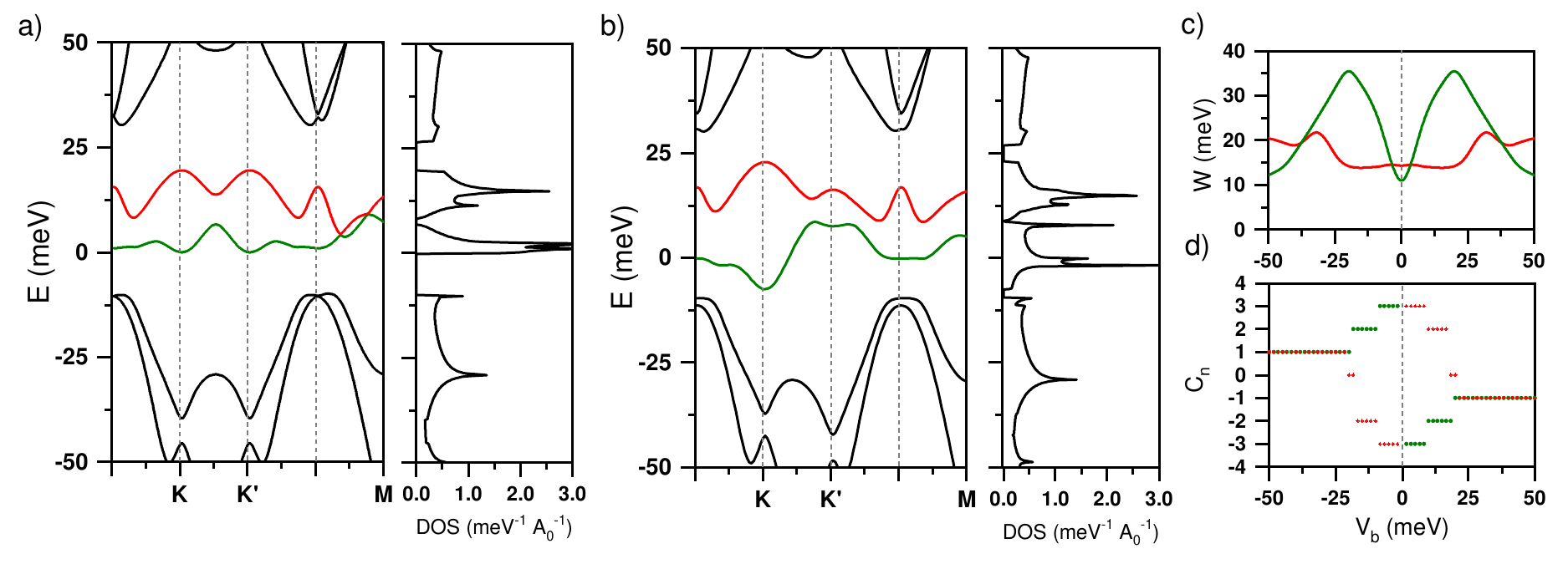} 
\par\end{centering}
\caption{Band structure of two twisted graphene bilayers with a twist angle of $\theta = 1.24^\circ$ and for a) $V_{b}=0$ we find two middle narrow bands. Along the $\Gamma M$ path there are two points where these bands cross. b) Band structure with an electrostatic potential of $V_{b}=5$ meV. In c) and d) we show the bandwidth and the Chern number of the middle bands as a function of the electrostatic potential, respectively. Each of the middle bands is labelled by the same color in all figures.  
\label{fig: FigTDBGbands}}
\end{figure*}

\section{Two twisted graphene bilayers} \label{sec: Two twisted graphene bilayers}

\subsection{Band structure}

A graphene bilayer twisted with respect to another graphene bilayer can be modelled using a continuum model, as defined in \cite{LopesDosSantos2007b}; related geometries have been studied in \cite{Cea2019a}. We will refer to this system as twisted double-bilayer graphene (TDBG). This model describes the interlayer hoppings between the two neighboring graphene monolayers which are twisted with respect to each other, while Eq.~\eqref{eq: BilMatrix} describes interlayer tunneling within each bilayer. The overall stacking is $ABA'B'$ as in Ref. \cite{Koshino2019a}. 

Recent experiments on TDBG~\cite{Shen2019, Cetal19} reported the observation of insulating phases at integer fillings, which clearly indicate the important role of many-body interactions~\cite{Shen2020,HeSym2020,Haddadi2020,Chandra2020}. Remarkably, the sensitivity of the insulating gap to an in-plane magnetic field may indicate that the insulating states are ferromagnetic insulators, in contrast to the Mott insulator states that are the most likely candidates in TBG~\cite{Shen2019, Cetal19}. The ferromagnetic fluctuations may provide an electronic additional mechanism for the formation of the superconducting order. Recent theoretical studies emphasize the sensitivity of the band structure to an out-of-plane electric field, which may easily open gaps between very narrow mini-bands, and the behavior of the Chern number upon varying the stacking arrangements~\cite{Chebrolu2019a,Koshino2019a,Wang2021OChern,XiaTDBG2020}.  We consider a twist angle $\theta = 1.24^\circ$, which has been reported to be favorable for superconductivity~\cite{Letal19}. In this case we find the narrowest bands near zero displacement field, while finite displacement fields produce wider bandwidths. Results are shown in Fig.~\ref{fig: FigTDBGbands} and in Fig.~\ref{fig: FigTDBGHar4}. The bands reported in Fig.~\ref{fig: FigTDBGbands} are in reasonable agreement with those in~\cite{Koshino2019a}. As in the two previous cases, and similar to related materials~\cite{goodwin2021flat}, even for a small dielectric constant the Hartree potential only slightly modifies the shape and bandwidth of the narrow bands. 

\subsection{Self-consistent Hartree interaction}

\begin{figure} 
\begin{centering}
\includegraphics[scale=0.57]{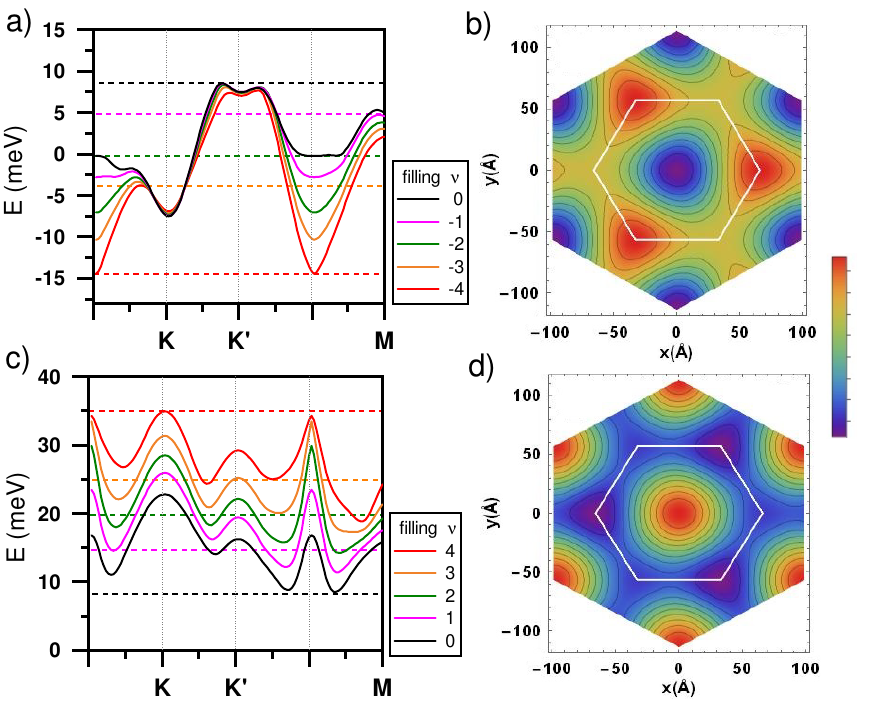} 
\par\end{centering}
\caption{
Self-consistent bands for TDBG in the presence of a Hartree potential. We consider a twist angle $\theta =1.24^\circ$, $V_{b}=5$~meV and dielectric constant $\epsilon=4$. In a) and c) we plot the lower and upper narrow band with negative and positive filling fractions, respectively. Panels b) and d) display the real space distribution of the Hartree potential for a filling fraction  $\nu=-2$ and $\nu=2$, respectively. The white hexagon shows the real-space unit cell. The horizontal dashed lines in a) and c) are the corresponding Fermi energies.
\label{fig: FigTDBGHar4}}
\end{figure}

\begin{figure*}     
\begin{centering}

\includegraphics[scale=0.3]{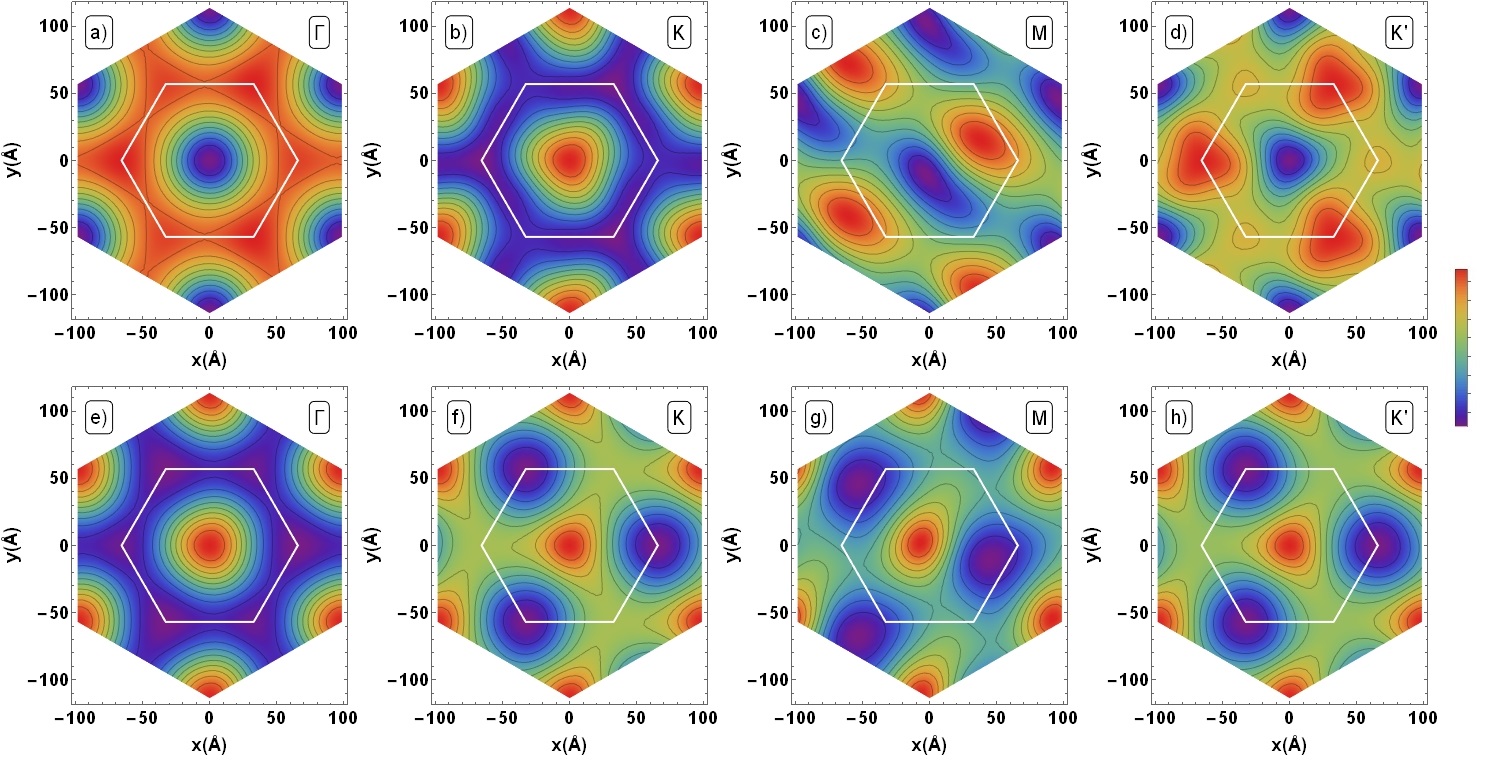}
\par\end{centering}
\centering{}\caption{Charge densities at the high symmetry points within the mBZ for $V_{b}=5$ meV, $\epsilon =4$ and a twist angle of $\theta =1.24$. The corresponding symmetry point is indicated in each panel. The top (bottom) row of figures is the charge density for the points of the  bands with $\nu=-2$ ($\nu=2$) in Fig.~\ref{fig: FigTDBGHar4}. The coloring corresponds to purple for the minimum and red for the maximum charge density. \label{fig: FigTDBGDens}}
\end{figure*}

Similar to TBG, we find that the dominant contribution to the Hartree interaction comes from the first star of reciprocal lattice vectors. Independently of the filling, the Fourier components of the Hartree potential (Eq.~\eqref{eq: rho components}) are real numbers with a small imaginary component. In Fig.~\ref{fig: FigTDBGDens} we show the charge density calculated in different points of the mBZ, as specified in the labels. The top (bottom) row refers to the charge density of states corresponding to the lower (upper) band in the middle of the spectrum, when the filling fraction is $\nu= -2$ ($\nu=2$). Similar to the previous cases of BG/hBN and TG/hBN, the effect of the Hartree potential is a band shifting depending on the shape of the charge density in the shifted region. For example, in the case of a negative  filling in Fig.~\ref{fig: FigTDBGHar4}a), there is a band shifting near the $\vec \Gamma$ point. In this region, the states experience the Hartree potential in Fig.~\ref{fig: FigTDBGHar4}b). Notice that states in the other regions, $\vec K$ or $\vec K'$, are insensitive to this potential. In the case of positive filling, states in all mBZ are subject to a similar Hartree potential as that in Fig.~\ref{fig: FigTDBGHar4}d) and the bands are uniformly shifted.

\section{Fock Interaction in all three systems} \label{sec: Fock Interaction}

We now consider the effect of the Fock potential in the three systems considered here. We follow the procedure described in Ref.~\cite{Cea2019,Cea2020c} by only considering those solutions where there is no spin or valley polarization. The matrix elements of the exchange potential $V_\text{F}$ are given by
\begin{equation}
v_{F,\boldsymbol{k}}\left(\boldsymbol G^{\prime}, \boldsymbol G\right)=-\sum_{\boldsymbol{q},l,\boldsymbol {G''}}\frac{V_{C}\left(\boldsymbol{q}-\boldsymbol{k}-\boldsymbol{G''}\right)}{\Omega}C_{q,l}\left(\boldsymbol G,\boldsymbol G^{\prime},\boldsymbol G''\right)
\label{eq: vfock}
\end{equation}
where $V_{C}(q)$ is the Fourier transform of the Coulomb potential and
\begin{equation}
C_{q,l}\left(\boldsymbol G,\boldsymbol G^{\prime},\boldsymbol{G''}\right)=\phi_{\boldsymbol{q},l}\left(\boldsymbol{G^{\prime}}+\boldsymbol{G''}\right)\phi_{\boldsymbol{q},l}^{*}\left(\boldsymbol{G}+\boldsymbol{G''}\right),
\label{eq: vfockcoup}
\end{equation}
is an overlap term. 
In Eq.~\eqref{eq: vfock}, $\Omega$ is the area of the real space unit cell and $l$ runs over all occupied states above a given threshold. Here we set this threshold to the lowest energy of the narrow bands (for details please refer to Ref.~\cite{Cea2020c}) and we only consider the effect at charge neutrality (CN), where the Fock term is largest. The matrix elements in Eq.~\eqref{eq: vfock} for $\boldsymbol G = \boldsymbol G'$ are real numbers and they contribute as an on-site momentum dependent term. The non-diagonal terms with $\boldsymbol G \neq \boldsymbol G'$ involve overlaps between the components of the wavefunctions. Our numerical analysis indicates that the diagonal on-site terms strongly dominates over the overlap terms. In TBG, the diagonal terms contribute to the broken $\mathcal{C}_{2}$ symmetry~\cite{Cea2020c}. In the systems considered here, this symmetry is already broken by the presence of the substrate or by a perpendicular displacement field.  

\begin{figure*}    
\begin{centering}
\includegraphics[scale=0.60]{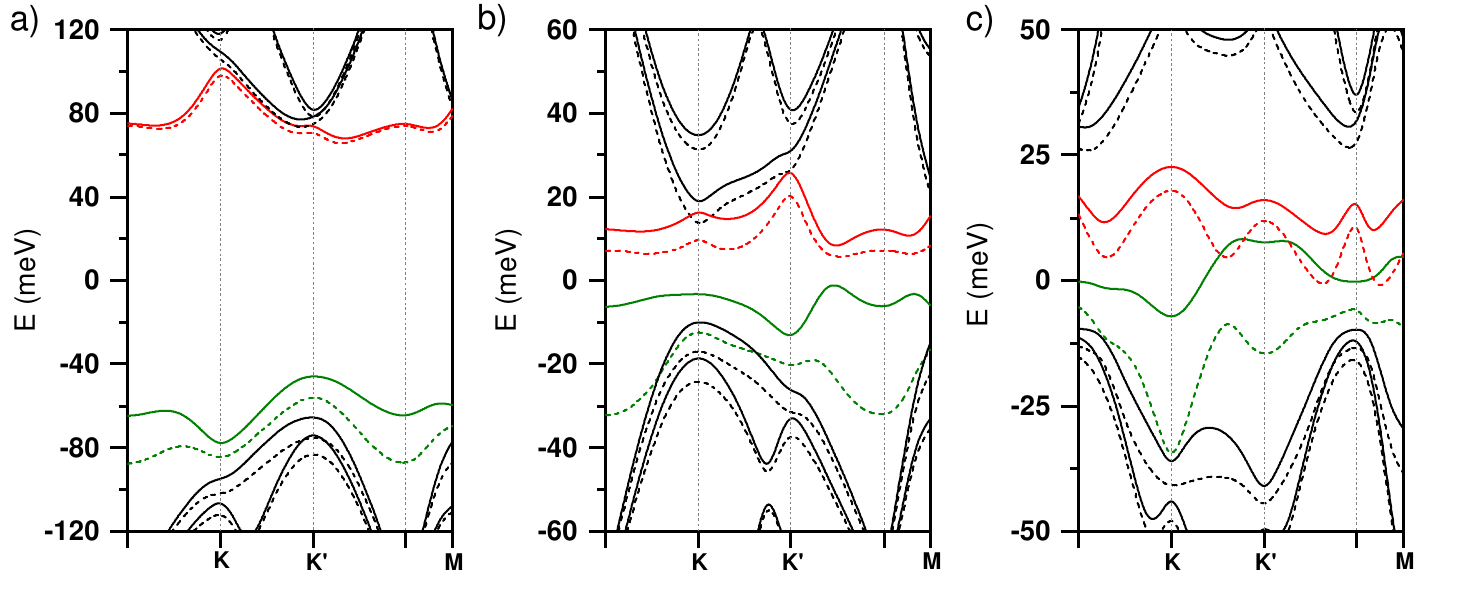}
\par\end{centering}
\caption{Band structure of a) BG/hBN, b) TG/hBN and c) TDBG with (dashed lines) and without (continuous lines) a self-consistent Fock interaction. Parameters in each panel are as in Fig.~\ref{fig: FigBGbands}b), Fig.~\ref{fig: FigTGbands}a) and Fig.~\ref{fig: FigTDBGbands}b), respectively.  The self-consistent solutions are obtained at charge neutrality where the Fock interaction is larger.    
\label{fig: FigFockAll}}
\end{figure*}

Figure~\ref{fig: FigFockAll} displays the band structure for a) BG/hBN, b) TG/hBN and c) TDBG. Continuous lines are the bands without interactions and dashed lines are the self-consistent solutions with a Fock interaction. In the case of BG/hBN in Fig.~\ref{fig: FigFockAll}a) we use the same parameters as in Fig.~\ref{fig: FigBGbands}b) where the bands are narrow. Here, the Fock interaction results in an almost constant shift of the lower state of the narrow band. In TG/hBN in Fig.~\ref{fig: FigFockAll}b) the distortion is stronger because the narrow band has a smaller bandwidth. The larger shift is around the $\Gamma$ point and is smaller at the mBZ boundary. The bandwidth of the occupied band is increased from $13$ to $21$~meV. In both BG/hBN and TG/hBN we find that the band shift due to interactions is larger at the center of the mBZ than at the boundaries. In the case of TDBG the effect is exactly the opposite, the band shift at the boundaries of the mBZ is larger than in the center, as shown in Fig.~\ref{fig: FigFockAll}c). 

In the three systems considered here, we have only taken into account matrix elements with $\boldsymbol G = \boldsymbol G'$ in Eq.~\ref{eq: vfock}. The non-diagonal terms are quite small and they can be safely neglected. In this situation, the Fock interaction behaves as a momentum dependent on-site contribution. In twisted bilayer graphene, on the other hand, this potential includes significant periodic contributions commensurate with the moir\'e lattice. The independence of the Fock potential on the moir\'e superlattice in the systems studied here implies that broken symmetry states induced by long range electrostatic interactions do not require a fine tuning of the effects of the substrate, which seems confirmed by recent experiments~\cite{zhou2021half,zhou2021superconductivity}.

Although we only focused our analysis at charge neutrality where the Fock interaction is large, the strong on-site potential can be captured by an effective on-site Hubbard interaction \cite{ZHU2018spiral,Zheng2020uf,RepellinFer2020,FengchengFerro2020,spethmann2021magnetic} and this may explain the Mott insulating states observed in TG/hBN~\cite{Chen2019,Chen2019a}, the ferroelectricity in BG/hBN~\cite{Zheng2020uf} and correlated effects in TDBG~\cite{Shen2020,HeSym2020,Haddadi2020,Chandra2020,Shen2019, Cetal19}. These effects are different from those in TBG where the Fock on-site terms are small and the overlap terms are large~\cite{Cea2019,Cea2020c}. Our results clearly indicates that there are significant differences between the narrow bands in TBG and related materials and narrow bands in other graphene configurations.

\section{Conclusions} \label{sec: Conclusions}
We have analyzed the leading effect of the electron-electron interactions, the long range Coulomb interaction treated within the Hartree-Fock approximation, in three types of stacks of graphene layers where very narrow bands emerge: i) untwisted AB bilayer graphene aligned with hBN, ii) trilayer ABC graphene aligned with hBN, and iii) two graphene bilayers twisted with respect to each other. In the three cases, on the application of an out of plane displacement field the bandwidth of bands near the neutrality point becomes comparable to or below 20 meV. In the absence of interactions, the bands for each of the two valleys show a complex pattern of crossings, Berry curvature, and Chern numbers tunable by perpendicular electrostatic potentials. The Hartree potential leads to small changes in the shape and width of the bands, even when the screening is small and the band is very narrow. This behavior implies that the long range electrostatic interaction plays a significantly different role in these systems than in twisted bilayer graphene, where similar calculations lead to changes in the bandwidth of order of $e^2/(\epsilon L)\sim 10 - 20$ meV, where $\epsilon \sim 4 - 10$ is the dielectric constant of the environment and $L$ is is the length of the unit vector of the moir\'e lattice. We found that the Fock terms contribute effectively as a momentum dependent on-site term and strongly distort the band structure. An exhaustive analysis of the Fock term in moiré superlattices has been provided so far in the Refs.~\cite{Xie2020a, Cea2020c}, for the case of magic-angle twisted bilayer graphene (MATBG). Depending on the filling of the conduction band the Fock term may break the $C_{2}$ symmetry of the MATBG, producing a gap between the Dirac points of the two constitutive graphene layers. Even though the magnitude of these gaps is of a few meVs, comparable to the bandwidth of the MATBG close to the charge neutrality point (CNP), it is sufficient to gap the Fermi surface and lead to \text{\textquotedblleft weak\textquotedblright} insulating states at integer fillings of the conduction band. The electronic structure of the systems considered in this work differs from that of the MATBG in the following aspects: i) the $C_{2}$ symmetry is generally not preserved even in the non-interacting case, which is in most cases is caused by the hBN substrate; ii) the non-interacting bandwidth close to the charge neutrality is at least one order of magnitude larger than that of the MATBG. As a consequence the Hartree term barely affects the states while the on-site Fock terms strongly affect the electronic band structure. This assumption is opposite with the recent finding of Ref.~\cite{Cea2020a}, that the Fock term is negligible in MATBG on a substrate of hBN. The results presented here suggest that there are significant differences between the narrow bands in twisted graphene bilayers and narrow bands in other graphene configurations; a similar analysis, derived from experimental findings, can be found in Ref.~\cite{HeSym2020}.

\section*{Acknowledgements}
This work was supported by funding from the European Commision, under the Graphene Flagship, Core 3, grant number 881603, and by the grants NMAT2D (Comunidad de Madrid, Spain),  SprQuMat and SEV-2016-0686, (Ministerio de Ciencia e Innovación, Spain).


\setcounter{equation}{0}
\setcounter{figure}{0}
\renewcommand{\theequation}{A\arabic{equation}}
\renewcommand{\thefigure}{A\arabic{figure}}

\section*{Appendix A: Low energy model for TDBG} \label{App: Appendix A: Low energy models}

\begin{figure}                     
\includegraphics[scale=0.20]{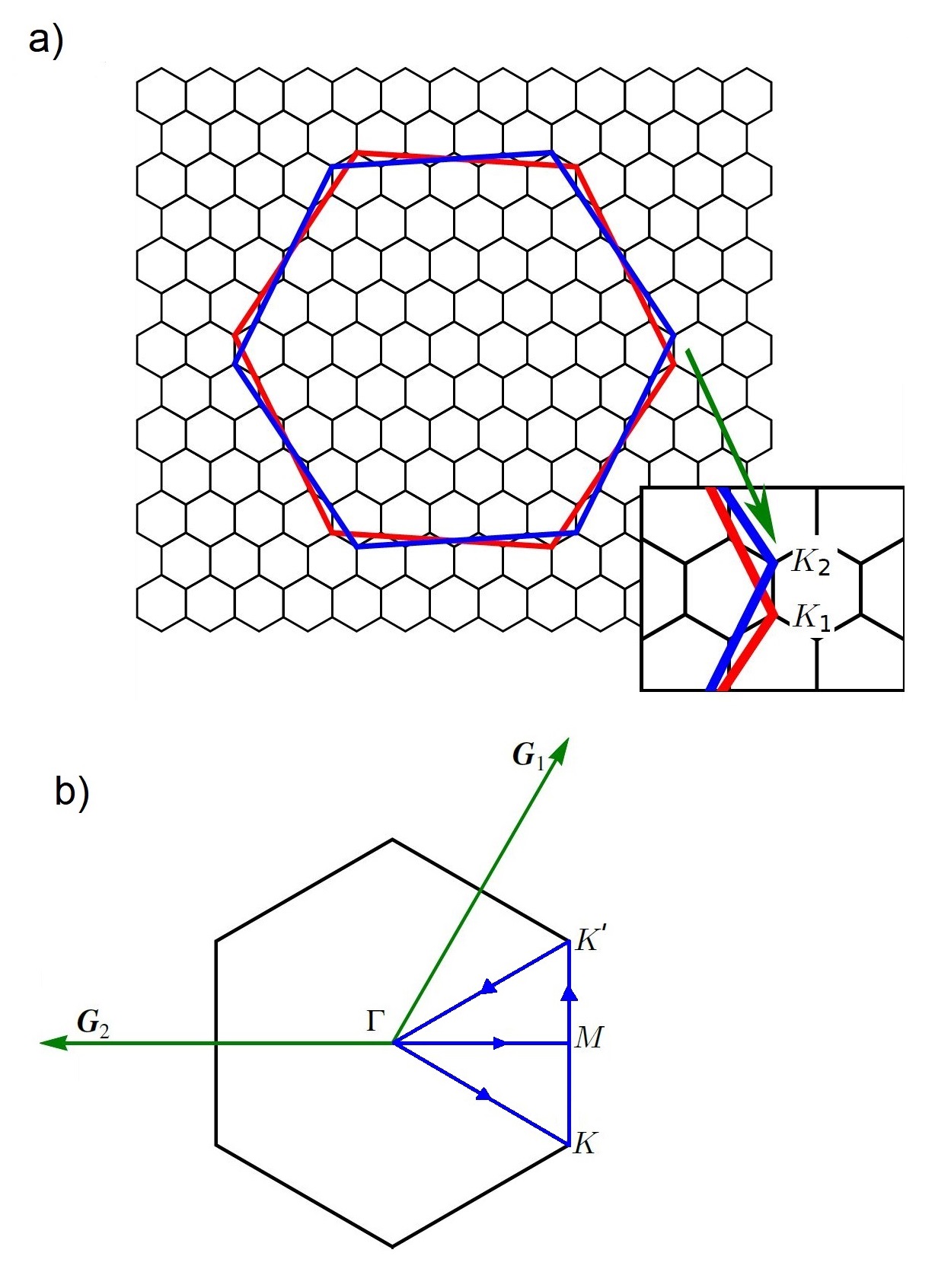}
\caption{ (a) Folding of the BZs of the twisted bilayers of graphene.
The BZ of the bottom bilayer (red hexagon) is rotated by $-\theta/2$, while that of the top bilayer 2 (blue hexagon) by $\theta/2$. The small black hexagons represent the moir\'e BZs forming the reciprocal moir\'e lattice. In the inset: $K_{1,2}$ are the Dirac points of the twisted bilayers, which identify the corners of the mBZ. (b) TDBG moir\'e Brillouin Zone. $\vec{G}_{1,2}$ are the two reciprocal lattice vectors.}
\label{fig: TDBGmBZ}
\end{figure}

In the case of TDBG, the relative twist between two Bernal stacks of bilayer graphene leads to the appearance of a moir\'e pattern, identical to that occurring in TBG. The size of the supercell, $L_m=a/2\sin\left(\theta/2\right)$, dramatically increases with the twist angle, $\theta$, $a=0.246$ nm being the lattice constant of graphene. Thus, for $\theta\sim 1.24^\circ$, $L_m\sim 11.36$ nm.  We describe the low-energy band structure of the TDBG within the continuum model introduced in the Refs.~\cite{LopesDosSantos2007b,Bistritzer2011} for the case of the TBG, and generalized in Refs.~\cite{Koshino2019a,Chebrolu2019a}. This model is meaningful for sufficiently small angles, so that an approximately commensurate structure can be defined for any twist. The moir\'e mini-BZ, resulting from the folding of the two BZs of each bilayer, has the two reciprocal lattice vectors:
\begin{equation}
\vec{G}_1=2\pi(1/\sqrt{3},1)/L\text{ and } \vec{G}_2=4\pi(-1/\sqrt{3},0)/L,
\end{equation}
 shown in green in Fig.~\ref{fig: TDBGmBZ}(b). For small twists, the coupling between the two valleys at $K=(4\pi/3a)(1,0)$ and $-K$ of the unrotated bilayer can be safely neglected, as the interlayer hopping has a long wavelength modulation. Then, for the sake of simplicity, in what follows we focus only on the $K$-valleys of each bilayer. The Hamiltonian of the TDBG is represented by the $8\times 8$ matrix:
\begin{equation}
H_{TDBG}=\begin{pmatrix}
H_0(\vec{k}_1)&g(\vec{k}_1)&0&0&&\\
g^\dagger (\vec{k}_1) &H'_0(\vec{k}_1)&U&0&\\
0&U^\dagger& H_0(\vec{k}_2)&g(\vec{k}_2)\\
0&0&g^\dagger (\vec{k}_2) &H'_0(\vec{k}_2)
\end{pmatrix}
\label{eq: HamilTDBG}
\end{equation}
acting on the Nambu spinor $\Psi^T=\left(\psi_{A1},\psi_{B1},\psi_{A2},\psi_{B2},\psi_{A3},\psi_{B3},\psi_{A4},\psi_{B4}\right)$, whose entries are labels by the sub-lattice ($A/B$) and layer ($1,\dots 4$) indices. Here we defined $\vec{k}_l=R\left((-)^{l-1}\theta/2\right)\left( \vec{k}-\vec{K}_l \right)$, $R(\theta)$ being the $2\times 2$ matrix describing the counter-clock-wise rotation of the angle $\theta$, and:
\begin{align}
H_0(\vec{k})&=\begin{pmatrix}
    0&\hbar v_F k_-\\
    \hbar v_F k_+& \Delta
    \end{pmatrix} \notag \\
H'_0(\vec{k})&=\begin{pmatrix}
    \Delta&\hbar v_F k_-\\
    \hbar v_F k_+& 0
    \end{pmatrix} \\
g(\vec{k})&=
    \begin{pmatrix}
   -\hbar v_4k_-&-\hbar v_3k_+\\   \notag
    \gamma_1& -\hbar v_4 k_-
    \end{pmatrix}
\end{align}
where $k_{\mp}=k_x\mp ik_y$, $v_i=(\sqrt{3}/2)\gamma_i a/\hbar$, $\gamma_1=0.4$ eV, $\gamma_3=0.32$ eV, $\gamma_4=0.044$eV, $\Delta=0.05$eV, $\hbar v_F/a=2.1354$eV (see e.g. the Ref. \cite{Koshino2019a}). The $2\times 2$ matrix $U$ describes the moir\'e potential generated by the hopping amplitude between $p_z$ orbitals localized at opposite layers of the two twisted surfaces. In real space, $U(\vec{r})$ is a periodic funcion in the moir\'e unit cell. In the limit of small angles, its leading harmonic expansion is determined by only three reciprocal lattice vectors~\cite{LopesDosSantos2007b}:
$U(\vec{r})=U(0)+U\left(-\vec{G}_1\right)e^{-i\vec{G}_1\cdot\vec{r}}+
U\left(-\vec{G}_1-\vec{G}_2\right)e^{-i\left(\vec{G}_1+\vec{G}_2\right)\cdot\vec{r}}$,
where the amplitudes $U\left(\vec{G}\right)$ are given by:
\begin{align}
U(0)&=\begin{pmatrix}g_1&g_2\\g_2&g_1\end{pmatrix}\\   \notag
U\left(-\vec{G}_1\right)&=\begin{pmatrix}g_1&g_2e^{-2i\pi/3}\\g_2e^{2i\pi/3}&g_1\end{pmatrix}\\  
U\left(-\vec{G}_1-\vec{G}_2\right)&=U\left(-\vec{G}_1\right)^*   \notag
\end{align}
In the following we adopt the parametrization of the TBG given in the Ref.\cite{Koshino2018a}: $g_1=0.0797$ eV and $g_2=0.0975$ eV. The difference between $g_1$ and $g_2$ accounts for the corrugation effects where the interlayer distance is minimum at the $AB/BA$ spots and maximum at $AA$ ones, or can be seen as a model of a more complete treatment of lattice relaxation \cite{Guinea2019a}. The difference in electrostatic energy between the adjacent bands is given by~\cite{Koshino2019a}  

\begin{equation}
V_{E}=\frac{1}{2}V_{b}\left(\begin{array}{cccc}
3I_{2} & 0 & 0 & 0\\
0 & I_{2} & 0 & 0\\
0 & 0 & -I_{2} & 0\\
0 & 0 & 0 & -3I_{2}
\end{array}\right),
\label{eq: TBGVias}
\end{equation}
where $I_{2}$ is a $2\times2$ identity matrix and $V_{b}$ an (out of plane) external electrostatic potential. The long-range Coulomb interaction, is introduced as a self-consistent Hartree potential, $V_\text{H}$. 

\section*{Appendix B: Self-consistent Hartree interaction} \label{App: Appendix B: Hartree}

The matrix elements of the self-consistent Hartree potential are given by \cite{Guinea2018a,Cea2019}
\begin{equation}
\rho_{H}(\vec G)=4V_{C}\left(\vec G\right) \int \frac{d^{2}\boldsymbol{k}}{V_{mBZ}}\sum_{\vec G^{\prime},l}\phi_{k,l}^{\dag}\left(\vec G^{\prime}\right)\phi_{k,l}\left(\vec G+\vec G^{\prime}\right),\label{eq: rho components} 
\end{equation}
with $V_{C}\left(\vec G\right)=2\pi e^{2}/\epsilon \left|\vec G\right|$  the Fourier transform of the Coulomb potential evaluated at $\vec G$, $V_{mBZ}$ is the area of the mBZ and the factor 4 takes into account spin/valley degeneracy. The parameter $l$ is a band index resulting from the diagonalization of the full Hamiltonian (Eq.~(\ref{eq: Main Hamiltonian}) for (TG)BG/hBN and Eq.~(\ref{eq: HamilTDBG}) for TDBG) and $\phi_{k,l}\left(\vec G \right)$ is the amplitude for an electron to occupy a state with momentum $\vec k+\vec G$.  The value of $\rho_{H}(\vec G)$ in Eq.~(\ref{eq: rho components}) depends on the extent of the wavefunctions in momentum space, hence we can write the Fourier expansion of the Hartree potential in real space as
\begin{equation}
V_\text{H}(\vec{r})=V_{0}\sum_{n}\rho_\text{H}(\vec G_{n})e^{i\vec G_{n}\cdot\vec{r}}
\label{eq: VhRealSpace}
\end{equation}
with $\rho_\text{H}(\vec G_{n})=\left|\rho_\text{H}(\vec G_{n})\right|e^{i\arg\left[\rho_\text{H}(\vec G_{n})\right]}$ a complex number with $-\pi<\arg\left[\rho_\text{H}(\vec G_{n})\right]<\pi$ and $V_{0}=e^{2}/\varepsilon L$  the effective Coulomb potential. Due to the triangular symmetry of the superlattice potential the above equation can be written as
\begin{equation}
V_\text{H}(\vec{r})=2V_{0}\sum_{n}\left|\rho_\text{H}(\vec G_{n})\right|\cos\left(\arg\left[\rho_\text{H}(\vec G_{n})\right]+\vec G_{n}\cdot\vec{r}\right).
\label{eq: FourierHartree}
\end{equation}
To solve the self-consistent Hartree Hamiltonian, the charge distribution is approximated as $\rho_{H}=\overline{\rho}_{H}+\delta\rho_{H}$ where $\overline{\rho}_{H}$ is a constant which takes into account the total density from all bands not included in the calculations~\cite{Guinea2018a}. The charge distribution is fixed by considering an homogeneous state at the CNP, this is $\rho_{H}=0$. Therefore, the  integral in Eq.~(\ref{eq: rho components}) is evaluated only over energy levels with $E_l\left(\vec k\right)$ between the CNP and Fermi level. The matrix elements of the Hartree potential $V_H$ in Eq. (\ref{eq: rho components}) depends implicitly on the filling fraction $\nu$ of the conduction band. For a fully filled valence and conduction bands we have $\nu=4$ and when they are both empty we have $\nu=-4$. 
For a given value of the filling, we calculate the miniband spectrum and wavefunctions with self-consistent diagonalization of the main Hamiltonian by considering a coupling up to 5 stars, corresponding to a total of $N=91$ vectors in the reciprocal space. 

\widetext
\onecolumngrid
\section*{Appendix C: Hartree Potential in coordinate space} \label{App: Appendix C: Hartree Real Space}

\begin{figure*}     
\begin{centering}
\includegraphics[scale=0.30]{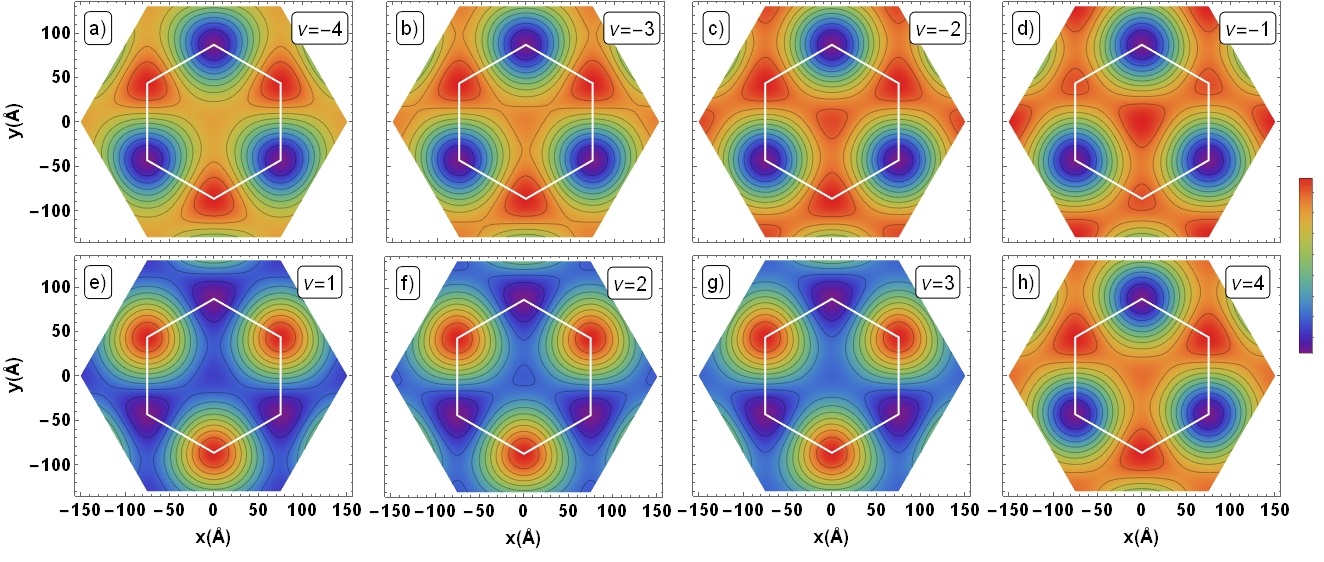}
\par\end{centering}
\centering{}\caption{Real space distribution of the Hartree potential for BG/hBN bands in Fig.~\ref{fig: FigBilHar4}. The coloring corresponds to purple for the minimum and red for the maximum Hartree potential. 
\label{fig: FigBGHartreeAll}}
\end{figure*}

\begin{figure*}     
\begin{centering}
\includegraphics[scale=0.27]{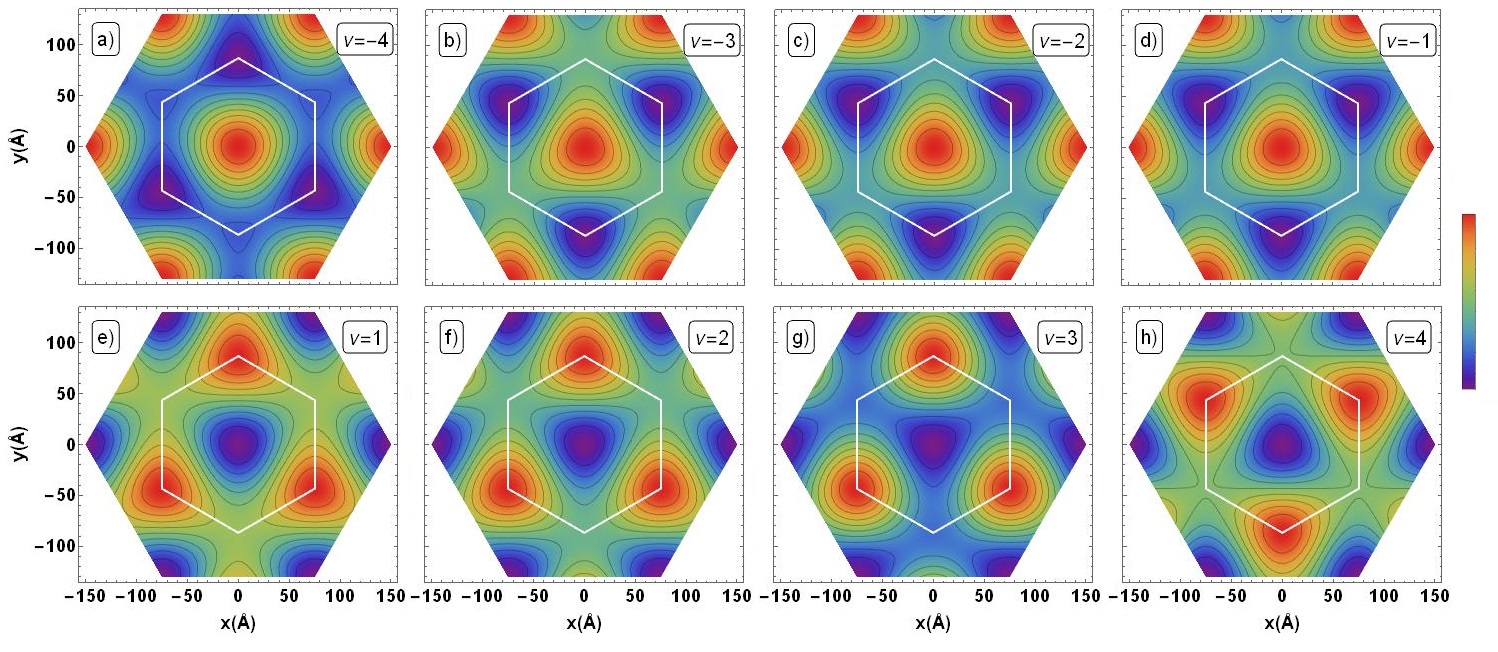}
\par\end{centering}
\centering{}\caption{Real space distribution of the Hartree potential for TG/hBN bands in Fig.~\ref{fig: FigTrilHar4}. The coloring corresponds to purple for the minimum and red for the maximum Hartree potential. 
\label{fig: FigTGHartreeAll}}
\end{figure*}

\begin{figure*}     
\begin{centering}
\includegraphics[scale=0.27]{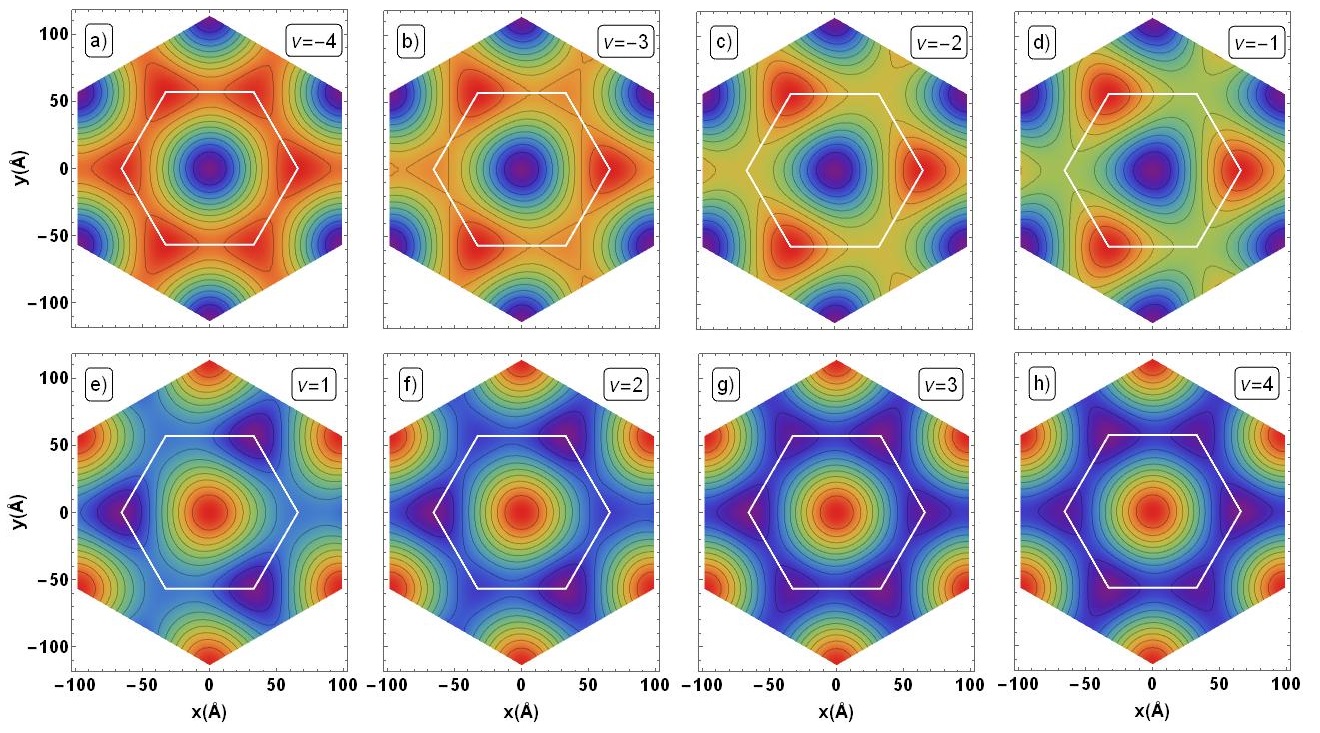}
\par\end{centering}
\centering{}\caption{Real space distribution of the Hartree potential for TDBG bands in Fig.~\ref{fig: FigTDBGHar4}. The coloring corresponds to purple for the minimum and red for the maximum Hartree potential. 
\label{fig: FigTDBGHartreeAll}}
\end{figure*}

We also study the evolution of the Hartree potential as a function of the filling of the flat bands. In Fig.~\ref{fig: FigBGHartreeAll} for BG/hBN and Fig.~\ref{fig: FigTGHartreeAll} for TG/hBN and Fig.~\ref{fig: FigTDBGHartreeAll} for TDBG  we show the the evolution of the Hartree potential in real space as a function of the filling fraction. The Hartree potential in Eq.~\ref{eq: VhRealSpace}, is written in terms of the Fourier coefficients $\rho_\text{H}$, which are complex quantities encoding the symmetric (real) and antisymmetric (imaginary) parts of the charge density. In TBG the coefficient $\rho_\text{H}$ is a real number because the charge density is centered at the unit cell origin~\cite{Rademaker2018a,Guinea2018a}. In all three cases considered here, the Fourier coefficient $\rho_\text{H}$ is a complex number and the charge density shows the complex patterns shown in  Fig.~\ref{fig: FigBGHartreeAll}, Fig.~\ref{fig: FigTGHartreeAll} and Fig.~\ref{fig: FigTDBGHartreeAll}.

\section*{Appendix D: Topological phases induced by Coulomb interactions}  \label{App: Appendix D: Topological Phases}

\begin{figure*}     
\begin{centering}
\includegraphics[scale=0.4]{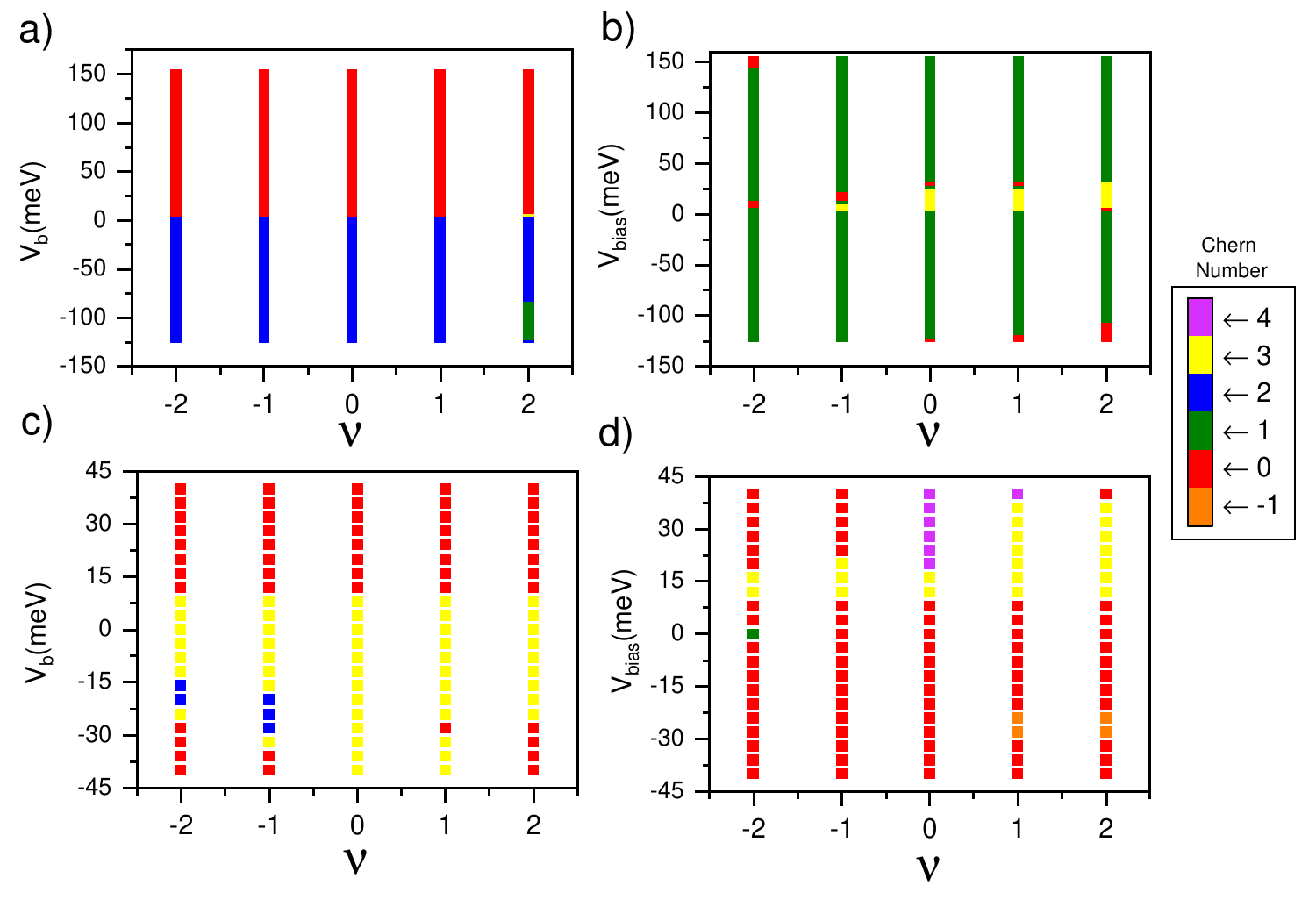}
\par\end{centering}
\centering{}\caption{Chern number as a function of the displacement field and integer filling fractions. for the a) lower and b) upper middle band of BG/hBN the Chern number changes from $0$ to $2$ depending on the filling and displacement field. In the c) lower and d) upper middle band of TG/hBN, the Chern number varies from $-1$ to $4$ in the considered range of fillings. 
\label{fig: Figtphases}}
\end{figure*}

In BG/hBN, non-local transport measurements have detected topological valley currents~\cite{Endo2019b}, indicating the existence of a non-trivial band topology. For TG/hBN, recent transport experiments~\cite{Chen2020} have found that Coulomb interaction driven topological phases are induced within the same device by modifying the filling fraction and the displacement field. In addition, by combining thermodynamic measurements, local and non-local transport measurements, it has also been shown that robust topologically non-trivial valley Chern insulators occur in TDBG~\cite{wang2021topological}. In the three systems considered here, we have found that there are topological phases induced by both displacement fields and the Coulomb interaction within the Hartree approximation. Figure~\ref{fig: Figtphases} displays the Hartree induced topological phases in the two narrow bands of BG/hBN (top panels) and in TG/hBN (bottom panels). In both cases, a large displacement field pushes the narrow bands close to the remote bands and even if the Hartree potential is small, is strong enough to induce topological phase transitions. In TDBG (not shown) the middle narrow bands are quite close to the remote bands. As shown in Fig.~\ref{fig: FigTDBGbands}d) an small displacement field induces topological phase transitions and therefore we also expect additional phases as a function of the filling.

\bibliographystyle{apsrev4-1}
\end{document}